\documentclass[fleqn,usenatbib]{mnras}

\usepackage{newtxtext,newtxmath}
\usepackage[T1]{fontenc}
\DeclareRobustCommand{\VAN}[3]{#2}
\let\VANthebibliography\thebibliography
\def\thebibliography{\DeclareRobustCommand{\VAN}[3]{##3}\VANthebibliography}
\usepackage{graphicx}
\usepackage{amsmath}
\usepackage{xcolor}
\usepackage{multirow}
\usepackage{subcaption}
\usepackage{makecell}
\usepackage{arydshln}

\date{}
\makeatletter
\def\@printed{
  \qquad\qquad\qquad Compiled using MNRAS \LaTeX\ style file v\@version}
\makeatother

\title[Interacting Dark Sectors and Upcoming Surveys]{Probing Interacting Dark Sectors with upcoming Post-Reionization and Galaxy Surveys}

\author[R. Shah, A. Dey, P. Mukherjee, S. Pal]{
Rahul Shah,$^{1}$\thanks{E-mail: rahul.shah.13.97@gmail.com}
Antara Dey, $^{1}$\thanks{E-mail: antaraaddey@gmail.com}
Purba Mukherjee,$^{1,2}$\thanks{E-mail: purba16@gmail.com}
and Supratik Pal$^{1}$\thanks{E-mail: supratik@isical.ac.in}\\
$^{1}$Physics and Applied Mathematics Unit, Indian Statistical Institute, 203 B.T. Road, Kolkata 700 108, India\\
$^{2}$Centre for Theoretical Physics, Jamia Millia Islamia, New Delhi 110025, India}
\begin{document}
\label{firstpage}
\pagerange{\pageref{firstpage}--\pageref{lastpage}}
\maketitle

\begin{abstract}
We investigate the constraining power of future post-reionization and galaxy surveys on possible interactions between dynamical dark energy and dark matter. The analysis focuses on the interaction strength and the dark energy equation of state parameters, in addition to the six standard cosmological parameters. Using fiducial values obtained from the current observational bounds (Planck 2018 + DESI DR2 + Pantheon+), mock datasets for upcoming 21-cm intensity mapping, galaxy clustering and cosmic shear observations from the SKA-mid, and for the upcoming large-scale survey from the Euclid mission, were generated. Subsequently, Markov chain Monte Carlo analyses combining current cosmological data with these mock datasets were performed to forecast parameter constraints. The results indicate that both SKA-mid and Euclid observations can significantly improve constraints on interacting dark sector parameters. In particular, the interaction strength and dark energy equation of state parameters can be constrained considerably tighter than current combined constraints from Planck 2018, DESI DR2 and Pantheon+. Comparing different probe combinations and survey configurations, it is found that SKA2 provides the tightest projected constraints, particularly on the interaction strength, while Euclid achieves a precision broadly comparable to that of SKA1. The results highlight the potential of these upcoming surveys to probe interactions within the dark sector.
\end{abstract}



\section{Introduction}\label{sec:introduction}

Modern cosmology is entering an era of increasingly precise observations driven by next-generation large-scale structure (LSS) and radio surveys. Upcoming experiments such as the Square Kilometre Array (SKA) \citep{Maartens:2015mra,Santos:2015gra} and the Euclid mission \citep{EUCLID:2011zbd,EuclidTheoryWorkingGroup:2012gxx} will probe an unprecedented volume of the observable Universe and map the distribution of matter across a wide range of redshifts. In particular, SKA will trace the distribution of neutral hydrogen using 21-cm intensity mapping (IM), while also enabling measurements of galaxy clustering (GC) and cosmic shear (CS). Euclid will provide high-precision optical and near-infrared observations of galaxy clustering and weak gravitational lensing. Together, these probes will allow detailed studies of the post-reionization Universe, where the large-scale distribution of neutral hydrogen and galaxies can be measured with high precision. These observations have the potential to significantly improve constraints on cosmological parameters and provide new tests of extensions to the ``vanilla" $\Lambda$ cold dark matter ($\Lambda$CDM) model.

Despite the success of the $\Lambda$CDM paradigm in explaining many cosmological observations, the physical nature of dark matter (DM) and dark energy (DE) remains unknown. In the standard framework, DM is assumed to be cold and non-relativistic, interacting only through gravity, while DE is described by a cosmological constant with an equation of state (EoS) $w=-1$. However, the increasing precision of cosmological measurements has revealed several tensions within this framework \citep{CosmoVerseNetwork:2025alb}. The most notable among these is the Hubble tension, which corresponds to a $\sim 7\sigma$ difference between the value of the Hubble constant inferred from observations of the cosmic microwave background (CMB) and the value measured in the local Universe using the cosmic distance ladder \citep{DiValentino:2021izs,H0DN:2025lyy}. In addition, discrepancies in the amplitude of matter clustering inferred from LSS observations, commonly quantified through the $S_8$ or $\sigma_{8,0}$ parameters, have also been reported \citep{DiValentino:2020vvd}. Furthermore, recent results from the Dark Energy Spectroscopic Instrument (DESI) suggest that the DE EoS may evolve with cosmic time, indicating a possible deviation from a cosmological constant \citep{DESI:2024mwx,DESI:2025zgx}. These developments have increased interest in exploring extensions to the $\Lambda$CDM model.

Among the proposed extensions, interacting dark matter dark energy (iDMDE) models have received considerable attention \citep{Wang:2016lxa}. In these scenarios, DM and DE can exchange energy \citep{Farrar:2003uw,Pan:2023mie,Lee:2006za,Lee:2009ji} or momentum \citep{Simpson:2010vh,Linton:2021cgd}. This interaction modifies both the expansion history of the Universe and the growth of cosmic structures. Such models were originally proposed as a possible explanation of the coincidence problem \citep{Amendola:1999er,Zimdahl:2001ar} and have since been studied extensively in many cosmological contexts. In particular, interactions between DM and DE have been explored as a possible way to alleviate the Hubble tension \citep{DiValentino:2017iww,Bhattacharyya:2018fwb,DiValentino:2019ffd,Lucca:2020zjb,Petri:2025swg,Yashiki:2025loj,Zhang:2025dwu,vanderWesthuizen:2025mnw,vanderWesthuizen:2025rip,Dai:2026pvx,Sahlu:2026bsa,Yadav:2026dhg}, as well as discrepancies in the amplitude of matter clustering \citep{Bhattacharyya:2018fwb,Sinha:2021tnr,Shah:2024rme,Shah:2025ayl,Tsedrik:2025cwc,Tsedrik:2025jdv}. Depending on the form of the interaction and the behaviour of the DE EoS, these models can produce distinctive signatures in LSS observables \citep{Silva:2025bnn,Wolney:2026zbc}. Future high-precision surveys therefore offer a promising way to test iDMDE scenarios \citep{Zhang:2021yof,Wang:2024vmw,Cruickshank:2025iig}. However, the ability of upcoming 21-cm intensity mapping and LSS surveys to constrain such models remains relatively unexplored, which motivates the present work.

This study investigates to what extent future LSS observations, including galaxy clustering \citep{Euclid:2020fjs} and cosmic shear measurements (from both 21-cm and optical/near-infrared surveys) \citep{Harrison:2016stv}, together with 21-cm intensity mapping observations \citep{Sprenger:2018tdb,SKA:2018ckk}, can constrain iDMDE models involving dynamical DE and DM. Considered here is a framework in which dark energy is described by a parametrized dynamical EoS, which interacts with DM through an energy transfer term proportional to the dark energy density. While interactions can also involve momentum transfer, we do not consider it in this work. This type of interaction has been widely studied in the literature and can lead to viable cosmological behaviour \citep{Bamba:2012cp,Wang:2016lxa,DiValentino:2017iww,DiValentino:2019ffd,Yang:2019uog,Lucca:2020zjb,Mukhopadhyay:2020bml,Liu:2022hpz,Hoerning:2023hks,Shah:2024rme,Li:2024qso,Goswami:2025uih,Li:2025muv,Li:2025owk,Li:2025ula,Shah:2025ayl,Silva:2025bnn,Silva:2025hxw,vanderWesthuizen:2025vcb,Yan:2025iga,Yang:2025jea,Li:2026xaz,Figueruelo:2026eis}. In addition to the standard six cosmological parameters, this analysis focuses on the interaction strength, and on the parameters that describe the evolution of the DE EoS.

The interaction strength affects the growth of matter perturbations and is therefore primarily constrained by probes sensitive to matter distribution and its evolution, such as galaxy clustering and 21-cm intensity mapping. In contrast, the parameters describing the dark energy EoS govern the expansion history and influence both distance measures and the growth of structure \citep{Linder:2003dr}, making them particularly sensitive to cosmic shear and geometrical measurements from galaxy surveys. The background density parameters are affected differently: the cold dark matter density, $\Omega_c$, directly influences the growth and amplitude of matter perturbations and is therefore strongly constrained by clustering and lensing observations, while the baryon density, $\Omega_b$, has a comparatively weaker impact on late-time observables and remains primarily constrained by CMB measurements.

To assess the sensitivity of upcoming surveys, forecast analyses were performed using 21-cm intensity mapping observations from the SKA-mid \citep{Maartens:2015mra}, together with galaxy clustering \citep{Euclid:2020fjs} and cosmic shear measurements \citep{Harrison:2016stv} expected from both SKA-mid and the Euclid \citep{Audren:2012vy,Euclid:2021qvm} mission. Mock datasets were generated using fiducial constraints obtained with current observational datasets. Instead of adopting $\Lambda$CDM as the fiducial model in all cases, the mock observations were generated separately for each cosmological scenario considered in this work. Moreover, both conservative and realistic treatments of non-linear scales were considered. This allows a systematic study of the constraining power of different probe combinations and survey configurations.

The main goals of this work are as follows. The first is to study the ability of future post-reionization and galaxy surveys, including 21-cm intensity mapping, galaxy clustering and cosmic shear, to constrain iDMDE models in which dark energy has a dynamical EoS. Second, to compare the expected performance of different survey configurations of SKA-mid, as well as that from Euclid, and to evaluate how combining these probes with current data improves constraints on the interaction strength and dark energy parameters. This is done by exploring how each individual parameter is getting constrained and investigating which ones are expected to be affected more by these upcoming missions. These analyses enable the quantification of the potential of upcoming surveys to test interactions within the dark sector.

The plan of the paper is as follows. Section \ref{sec:theory} describes the iDMDE framework and outlines the cosmological models used in this work. Section \ref{sec:data_methods} presents the forecasting methodology, the survey specifications, and the observables used in the analysis. The resulting constraints and their implications are discussed in Section \ref{sec:results}. Finally, the summarized findings and concluding remarks are in Section \ref{sec:conclusion}.


\section{Interacting Two-Fluid Set-Up with CDM and Dynamical DE}\label{sec:theory}

A phenomenological framework in which DM and DE are allowed to interact through non-gravitational processes is considered. The formalism adopted here follows the iDMDE framework discussed in \citet{Shah:2024rme}. Since the microscopic nature of both DM and DE remains unknown, constructing a fundamental Lagrangian description of their interaction is difficult \citep{Faraoni:2014vra}. A common approach is therefore to treat the two components as effective fluids that can exchange energy and momentum. This effective fluid description allows one to study the cosmological consequences of possible interactions without specifying the underlying particle physics model.

In the presence of a non-zero interaction, the energy-momentum tensors of the individual dark sector components are no longer conserved independently. Instead, their evolution satisfies,
\begin{equation}
    \nabla_\mu T^{\mu\nu}_{(k)} = \mathcal{Q}^\nu_{(k)},
\end{equation}
where the index $k$ denotes either DE (``de'') or DM (``dm''), and $\mathcal{Q}^\nu_{(k)}$ characterizes the exchange of energy and momentum between the two sectors. Conservation of the total energy-momentum tensor, dictated by the Bianchi identity, requires $\mathcal{Q}^\nu_{\mathrm{de}}=-\mathcal{Q}^\nu_{\mathrm{dm}}$. Here, we restrict to energy transfer interactions only, motivated by previous studies showing that such models can help alleviate cosmological tensions while providing a well-controlled and widely used framework; models including momentum exchange exist but are beyond the scope of this work. Consequently, the time components satisfy, $\mathcal{Q}^0_{\mathrm{de}}=-\mathcal{Q}^0_{\mathrm{dm}}=\mathcal{Q}$, while the spatial components vanish, $\mathcal{Q}^i_{(k)}=0$.

Assuming that DM is cold (CDM), \textit{i.e.} non-relativistic, the background continuity equations for the two components can be written as,
\begin{equation}
    \dot{\rho}_{\mathrm{dm}} + 3\mathcal{H}\rho_{\mathrm{dm}} = a^2 \mathcal{Q}^0_{\mathrm{dm}} = a\mathcal{Q},
\end{equation}
\begin{equation}
    \dot{\rho}_{\mathrm{de}} + 3\mathcal{H}(1 + w)\rho_{\mathrm{de}} = a^2 \mathcal{Q}^0_{\mathrm{de}} = -a\mathcal{Q},
\end{equation}
where a dot denotes a derivative with respect to conformal time, $\mathcal{H}=aH$ is the conformal Hubble parameter, and $w \equiv w(a)$ is the EoS parameter of DE. Here, conformal time $\tau$ is defined through $d\tau = dt/a$, where $a$ is the scale factor and $t$ is the cosmic time. The quantity $\mathcal{Q}$ specifies the rate of energy transfer between the two sectors. In the sign convention adopted here, $\mathcal{Q}>0$ corresponds to an energy flow from DE to DM, which is often considered in connection with the coincidence problem \citep{Olivares:2005tb, Pavon:2007gt}; in this case, the dark matter density dilutes more slowly than the standard $a^{-3}$ scaling, while dark energy decays faster, leading to a slower evolution of the ratio $\rho_{\mathrm{dm}}/\rho_{\mathrm{de}}$. This can help address the coincidence problem by keeping the two components comparable over a longer span of cosmic time, if sufficiently strong interactions are observationally supported.

Following a commonly used phenomenological prescription, the interaction term is assumed to be proportional to the dark energy density,
\begin{equation}
    \mathcal{Q} = H Q \rho_{\mathrm{de}},
\end{equation}
where $Q$ is a dimensionless constant coupling parameter that controls the strength of the interaction. The corresponding solutions of the background continuity equations are,
\begin{equation}
    \rho_{\mathrm{dm}}=\rho_{\mathrm{dm}, 0} \, a^{-3} \, + \rho_{\mathrm{de}, 0} \, a^{-3} \, Q \int_1^a \frac{a'^2 \, e^{\xi(a')}}{a'^Q} \, \mathrm{d} a',
\end{equation}
\begin{equation}
    \rho_{\mathrm{de}}=\rho_{\mathrm{de}, 0} \, \frac{e^{\xi(a)}}{a^Q},
\end{equation}
where $\xi(a)=\int_{a}^{1} \frac{3[1+w(a')]}{a'}\mathrm{d} a'$.

To study the growth of cosmic structures in this interacting scenario, perturbations in both the DM and DE components were considered. Working in the synchronous gauge and following the linear perturbation formalism gives, \citep{Ma:1995ey,Valiviita:2008iv,Gavela:2010tm,DaCosta:2014vsr,deCesare:2021wmk,Hoerning:2023hks},
\begin{equation}
    \dot{\delta}_{\mathrm{dm}}=-\theta_{\mathrm{dm}}-\frac{\dot{h}}{2}+\mathcal{H} Q \frac{\rho_{\mathrm{de}}}{\rho_{\mathrm{dm}}}\left(\delta_{\mathrm{de}}-\delta_{\mathrm{dm}}\right)+Q \frac{\rho_{\mathrm{de}}}{\rho_{\mathrm{dm}}}\left(\frac{k v_T}{3}+\frac{\dot{h}}{6}\right),
\end{equation}
\begin{equation}
    \dot{\theta}_{\mathrm{dm}}=-\mathcal{H} \theta_{\mathrm{dm}}\left(1+Q \frac{\rho_{\mathrm{de}}}{\rho_{\mathrm{dm}}}\right),
\end{equation}
\begin{align}
    \dot{\delta}_{\mathrm{de}}=&-(1+w)\left(\theta_{\mathrm{de}}+\frac{\dot{h}}{2}\right)- \frac{3\mathcal{H}\theta_{\mathrm{de}}}{k^2}\dot{w} -Q\left(\frac{k v_T}{3}+\frac{\dot{h}}{6}\right) \notag\\
    &-3 \mathcal{H}(c_s^2-w) \left[\delta_{\mathrm{de}}+\{3(1+w)+Q\} \frac{\mathcal{H} \theta_{\mathrm{de}}}{k^2}\right],
\end{align}
\begin{equation}
    \dot{\theta}_{\mathrm{de}}= -\mathcal{H} \left(1-3c_s^2\right) \theta_{\mathrm{de}} + \frac{Q\mathcal{H}\theta_{\mathrm{de}}}{1+w}\left(1+c_s^2\right) +\frac{k^2c_s^2}{1+w} \delta_{\mathrm{de}}.
\end{equation}
Here $\delta_{\mathrm{dm}}$ and $\delta_{\mathrm{de}}$ denote the density contrasts, while $\theta_{\mathrm{dm}}$ and $\theta_{\mathrm{de}}$ represent the velocity divergences of the two fluids. $\dot{h}$ denotes the derivative of the trace of the metric perturbation in the synchronous gauge. The background densities are $\rho_{\mathrm{dm}}$ and $\rho_{\mathrm{de}}$. The adiabatic sound speed of dark energy is taken to be equal to the EoS parameter $w$, while $c_s^2$ denotes the physical sound speed of DE perturbations. The quantity $v_T$ represents the centre-of-mass velocity of the total fluid, defined as $(1+w_T)v_T=\sum_i (1+w_i)\Omega_i v_i$. The additional terms appearing in the density contrast equations arise from perturbations in the Hubble expansion rate ($\delta H$) \citep{Hoerning:2023hks}. 

As representative dynamical dark energy models, two commonly used parametrizations of the DE EoS have been considered. The first is the Chevallier-Polarski-Linder (CPL) form, $w(a) = w_0 + w_a (1-a)$ \citep{Chevallier:2000qy,Linder:2002et}, and the second is the Jassal-Bagla-Padmanabhan (JBP) form, $w(a) = w_0 + w_a a(1-a)$ \citep{Jassal:2005qc}, where $w_0$ is the present day value, and $w_a$ is the evolution parameter. Notably, if one constructs a class of CPL-like parametrizations where the redshift-dependence scales as $z/(1 + z)^p$, with $p$ being a natural number, then the JBP parametrization is the second in this family with $p = 2$ after CPL. Consequently, JBP enables a steeper evolution of the EoS, while keeping the same value today and in the far past. These parametrizations provide simple and flexible descriptions of evolving dark energy and are widely used in studies of dynamical dark energy \citep{Copeland:2006wr,Li:2012dt}. They also allow a direct comparison with previous analyses of these iDMDE models \citep{Shah:2024rme,Shah:2025ayl}. In addition, recent observational hints of a possible evolution of the DE EoS from the Dark Energy Spectroscopic Instrument (DESI) \citep{DESI:2024mwx} further motivate the exploration of such parametrizations beyond the cosmological constant scenario.


\section{Datasets and Forecast Methodology} \label{sec:data_methods}

\subsection{Fiducial Cosmology and Mock Data Generation}

\begin{table*}
    \resizebox{1.0\textwidth}{!}{\renewcommand{\arraystretch}{1.2} \setlength{\tabcolsep}{20 pt}
    \begin{tabular}{c c c c c}
        \hline
        \textbf{Parameters} & \textbf{i-CPL Phantom} & \textbf{i-CPL Non-Phantom} & \textbf{i-JBP Phantom} & \textbf{i-JBP Non-Phantom} \\ 
        \hline
        \boldmath${100\Omega_b}{h^2}$ & $2.248$ & $2.256$ & $2.252$ & $2.251$ \\
        \boldmath${\Omega_c}{h^2}$ & $0.164$ & $0.037$ & $0.163$ & $0.144$ \\
        \boldmath$100{\theta_s}$ & $1.04188$ & $1.04200$ & $1.04207$ & $1.04187$ \\
        \boldmath${\ln{\left({10^{10}A_s}\right)}}$ & $3.031$ & $3.052$ & $3.045$ & $3.041$ \\
        \boldmath$n_s$ & $0.9683$ & $0.9715$ & $0.9713$ & $0.9698$ \\
        \boldmath${\tau}$ & $0.0496$ & $0.0580$ & $0.0575$ & $0.0531$ \\
        \boldmath$Q$ & $0.482$ & $-0.60$ & $0.48$ & $0.26$ \\
        \boldmath$w_0$ & $-1.016$ & $-0.730$ & $-1.020$ & $-0.863$ \\
        \boldmath$w_a$ & $-0.477$ & $-0.256$ & $-0.783$ & $-1.276$ \\
        \hline
        \boldmath$H_0$ & $67.77$ & $67.88$ & $67.98$ & $67.75$ \\
        \boldmath$\Omega_{m0}$ & $0.407$ & $0.132$ & $0.404$ & $0.364$ \\
        \boldmath$\sigma_{8,0}$ & $0.628$ & $1.784$ & $0.630$ & $0.691$ \\
        \boldmath$S_8$ & $0.731$ & $1.183$ & $0.731$ & $0.761$ \\
       \hline
         {\boldmath $\chi^2_{\rm min}$ } & 4201 & 4206 & 4204 & 4203 \\
        \hline
    \end{tabular}
    }
    \caption{The best-fitting constraints obtained for interacting CPL (i-CPL) and interacting JBP (i-JBP) using combined Planck 2018 + DESI DR2 BAO + Pantheon+ datasets \citep{Shah:2025ayl}.}
    \label{tab:bestfit}
\end{table*}

The forecasting strategy proceeds in two stages. First, mock observational data corresponding to each cosmological scenario considered in this work were constructed. The fiducial parameter values used to generate these mock datasets were obtained from the latest constraints derived from currently available cosmological observations. Specifically, the models were constrained using a combination of CMB from Planck 2018 TTTEEE + low-$\ell$ + low-E + lensing data \citep{Planck:2018lbu,Planck:2018vyg,Planck:2019nip}, DESI DR2 baryon acoustic oscillation (BAO) distance measurements reported in table IV of \citet{DESI:2025zgx} and type-Ia supernova (SNIa) observations from the Pantheon+ compilation \citep{Scolnic:2021amr}. For this purpose a modified version of \texttt{CLASS} \citep{Lesgourgues:2011re,Blas:2011rf} was employed, which is based on the iDMDE implementations of \citet{Lucca:2020zjb} and \citet{Hoerning:2023hks}, together with the Markov chain Monte Carlo (MCMC) sampler \texttt{MontePython} \citep{Audren:2012wb,Brinckmann:2018cvx}.

Rather than adopting the $\Lambda$CDM model as the fiducial cosmology in all cases, the mock observations were generated separately for each iDMDE scenario considered in this work \citep{Shah:2023rqb}. For each model, the best-fitting parameter values were chosen (Table \ref{tab:bestfit}), consistent with current observational constraints within that specific cosmological framework. The full constraints are reproduced from \citet{Shah:2025ayl} in Table \ref{tab:current} of Appendix \ref{appendix}. This approach ensures that the forecasted sensitivities of future surveys are evaluated in a manner consistent with the underlying model assumptions. The mock datasets were generated using the corresponding future-survey likelihoods implemented within \texttt{MontePython} \citep{Sprenger:2018tdb} as listed in Table \ref{tab:likes}.

The fiducial parameter vector considered in this analysis is $\{\Omega_{\rm b} h^2,\,\Omega_{\rm c}h^2,\,100\theta_{s},\,\ln\left(10^{10}A_{s}\right),\,n_s,\,\tau,\,Q,\,w_0,\,w_a\},$ where the first six parameters correspond to the standard cosmological parameters, while $Q$ represents the strength of the dark sector interaction and $(w_0, w_a)$ characterize the evolution of the DE EoS.

\subsection{Forecast Methodology}

\begin{table}
    \resizebox{0.5\textwidth}{!} {\renewcommand{\arraystretch}{1.5} \setlength{\tabcolsep}{25 pt}
    \begin{tabular}{c c}
        \hline
        \textbf{Likelihood} & \textbf{Description}\\
        \hline
        \textit{ska1\_IM\_band1} & SKA1 intensity mapping (IM) Band 1 \\
        \textit{ska1\_IM\_band2} & SKA1 intensity mapping (IM) Band 2 \\
        \textit{ska1\_pk} & SKA1 Galaxy Clustering (GC) \\
        \textit{ska1\_lensing} & SKA1 Cosmic Shear (CS) \\
        \textit{ska2\_pk} & SKA2 Galaxy Clustering (GC) \\
        \textit{ska2\_lensing} & SKA2 Cosmic Shear (CS) \\
        \textit{euclid\_pk} & Euclid Galaxy Clustering (GC) \\
        \textit{euclid\_lensing} & Euclid Cosmic Shear (CS) \\
        \hline
    \end{tabular}
    }
    \caption{Likelihoods utilized in this work as present in {\tt MontePython v3.6.1} \citep{Audren:2012wb,Brinckmann:2018cvx}, based on \citet{Sprenger:2018tdb}.}
    \label{tab:likes}
\end{table}

\begin{table}
    \resizebox{0.5\textwidth}{!} {\renewcommand{\arraystretch}{1.1} \setlength{\tabcolsep}{40 pt}
    \begin{tabular}{c c}
        \hline
        \textbf{Parameter}             & \textbf{Prior}\\
        \hline
        $\Omega_{\rm b} h^2$           & $[0.005,   0.1]$\\
        $\Omega_{\rm c} h^2$           & $[0.01,    0.99]$\\
        $100\theta_{s}$                & $[0.5,     10]$\\
        $\ln\left(10^{10}A_{s}\right)$ & $[1,       5]$\\
        $n_{s}$                        & $[0.5,     1.5]$\\
        $\tau$                         & $[0,       0.9]$\\
        $Q$                            & $[-2,      2]$\\
        $w_0$ (non-phantom)            & $[-1,      0]$\\
        $w_0$ (phantom)                & $[-3,      -1]$\\
        $w_a$                          & $[-2,      2]$\\
        \hline 
    \end{tabular}
    }
    \caption{Priors used for obtaining constraints on the cosmological parameters of interest.}
    \label{tab:priors}
\end{table}

The iDMDE models described in Section \ref{sec:theory} were implemented in the Boltzmann solver {\tt CLASS}\footnote{\url{https://github.com/lesgourg/class_public}} \citep{Lesgourgues:2011re,Blas:2011rf}, building upon the publicly available modified versions developed by \citet{Lucca:2020zjb} and \citet{Hoerning:2023hks}. Certain integrals arising from the dynamical EoS of dark energy were evaluated using Romberg integration. 

To avoid instabilities in the perturbation equations, special care was taken to ensure that the DE EoS $w(z)$ does not cross the phantom divide $w=-1$ during the evolution of the Universe. Within {\tt CLASS}, this was enforced by requiring that the quantity $(1+w(z))$ does not change sign between the infinite past and the present epoch $z=0$. In practice, the condition $\left(w(z=0)+1\right)\cdot\left(w(z)+1\right) > 0$ was imposed, which guaranteed that the model remains entirely within either the phantom or non-phantom regime and prevents unphysical perturbative instabilities associated with phantom crossing. The effects of the chosen dynamical dark energy parametrizations on the computation of the growth factor is accounted for as described by \citet{Linder:2003dr}. Throughout the analysis, the sound speed of dark energy perturbations is fixed to $c_s^2 = 1$, unless stated otherwise. 

Parameter constraints were obtained using the MCMC sampler {\tt MontePython}\footnote{\url{https://github.com/brinckmann/montepython_public}} \citep{Audren:2012wb,Brinckmann:2018cvx}. The MCMC approach allows a full exploration of the parameter space without assuming Gaussian likelihoods. While Fisher matrix forecasts provide a computationally efficient estimate of parameter uncertainties around a fiducial cosmology, they rely on Gaussian approximations and derivatives evaluated at the fiducial point. In contrast, Bayesian MCMC analyses can capture both Gaussian and non-Gaussian posterior distributions and are less sensitive to numerical instabilities that may arise in derivative-based methods. Hence, MCMC analyses were adopted to obtain the final forecasted parameter constraints. Moreover, no additional hard priors were imposed to explicitly exclude regions of parameter space that may lead to early-time perturbation instabilities, which are known to arise in interacting dark sector models. Instead, the full parameter space allowed by the model was explored, with such regions naturally disfavoured by the datasets considered.

The forecasts were performed using current cosmological datasets - CMB from Planck 2018 TTTEEE + low-$\ell$ + low-E + lensing \citep{Planck:2018lbu,Planck:2018vyg,Planck:2019nip}, BAO from DESI DR2 \citet{DESI:2025zgx} and SNIa from the Pantheon+ compilation \citep{Scolnic:2021amr} (henceforth referred to as CMB+BAO+SNIa or CBS) - in addition to mock datasets corresponding to SKA-mid and/or Euclid. All likelihood calculations were carried out within the {\tt MontePython} framework (as listed in Table \ref{tab:likes}). Throughout the analysis, a consistent set of priors (listed in Table \ref{tab:priors}) was used across all datasets and forecast combinations. The resulting parameter constraints and their implications are presented in the following section. Posterior distributions and contour plots are generated using {\tt GetDist}\footnote{\url{https://github.com/cmbant/getdist}} \citep{Lewis:2019xzd}.

\subsection{Forecasting with Future Surveys}

The second stage of the analysis closely follows the methodology of \citet{Dey:2023sxx}. The constraining power of upcoming LSS surveys using mock likelihoods corresponding to future observations was estimated. In particular, three primary cosmological probes expected from next-generation surveys were considered: 21-cm intensity mapping, galaxy clustering and cosmic shear. These observables provide complementary information on both the expansion history of the Universe and the growth of cosmic structures, making them powerful probes of iDMDE scenarios. The focus here is on the expected performance of SKA-mid and Euclid, both of which will map the large-scale distribution of matter over a wide range of redshifts.

\subsubsection{21-cm intensity mapping in the Post-Reionization Epoch}

After the epoch of reionization (EoR), the intergalactic medium becomes almost completely ionized for redshifts $z \lesssim 5$, with the remaining neutral hydrogen primarily residing within galaxies \citep{Mitra:2015yqa,Kulkarni:2018erh}. Observations of the redshifted 21-cm hyperfine transition of neutral hydrogen therefore provide a tracer of the large-scale matter distribution during the post-reionization epoch. One of the main observable quantities in 21-cm cosmology is the differential brightness temperature $\Delta T_{\rm b}$, which measures the difference between the hydrogen spin temperature $T_{\rm s}$ and the CMB temperature $T_{\nu}$. For radiation emitted at redshift $z$, the observed differential brightness temperature is given by \citep{Furlanetto:2006jb,Pritchard:2011xb},
\begin{equation}
    \Delta T_{\rm b} (z) = \frac{T_{\rm s}(z)-T_{\nu}(z)}{1+z}.
\end{equation}

In the post-reionization regime ($z \lesssim 5$) the spin temperature is typically much larger than the CMB temperature ($T_s \gg T_\nu$), allowing the signal to be expressed in terms of the mean brightness temperature,
\begin{equation}
    \overline{\Delta T_{\rm b}} (z) \approx 189 \left[\frac{H_{0}(1+z)^2}{H(z)}\right]\Omega_{\rm HI}(z) h \; {\rm mK},
\end{equation}
where $H_0=h\times100$ is the Hubble parameter $H(z)$ today and $\Omega_{\rm HI}= \rho_{\rm HI}(z)/\rho_{\rm c}$ is the neutral hydrogen density parameter. Fluctuations around the mean brightness temperature trace the underlying matter density fluctuations through,
\begin{equation}
\Delta T_{\rm b}(z) - \overline{\Delta T_{\rm b}}(z) = \overline{\Delta T_{\rm b}}(z)\,\delta_{\rm HI}(z) = \overline{\Delta T_{\rm b}}(z)\, b_{\rm HI}(z)\,\delta_{\rm m}(z)
\end{equation}
where $b_{\rm HI}$ is the neutral hydrogen bias and $\delta_{\rm m}$ denotes the matter density contrast. Consequently, the corresponding 21-cm power spectrum can be written as,
\begin{equation}
    P_{21}(k,z)= b_{\rm 21}^{2} P_{\rm m}(k,z) = (\overline{\Delta T_{\rm b}}\, b_{\rm HI})^{2} P_{\rm m}(k,z),
\end{equation}
where $P_m(k,z)$ denotes the linear matter power spectrum. 

\begin{table}
    \resizebox{0.5\textwidth}{!}{\renewcommand{\arraystretch}{1.2} \setlength{\tabcolsep}{5 pt}
    \begin{tabular}{c c c c c c c}
        \hline
        Experiment & $\nu_{\rm min}$ (MHz) & $\nu_{\rm max}$ (MHz) & $z_{\rm min}$ & $z_{\rm max}$ & $n_{bin}$ & $f_{sky}$ \\
        \hline
        SKA1 Band 1  & 400 & 1000 & 0.45 & 2.65 & 4 & 0.58  \\
        SKA1 Band 2 & 1000 & 1421 & 0.05 & 0.45 & 22 & 0.58  \\
        SKA1  & 950 & 1760 & 0.0 & 0.5 & 5 & 0.1212\\
        SKA2  & 470 & 1290 & 0.1 & 2.0 & 19 & 0.7272  \\
        Euclid & - & - & 0.7 & 2.0 & 13 & 0.3636\\        
        \hline
    \end{tabular}
    }
    \caption{Instrumental specifications adopted for SKA intensity mapping observations with SKA1 Band~1 and SKA1 Band~2 \citep{Olivari:2017bfv}, as well as for galaxy clustering and cosmic shear observations with SKA1, SKA2 \citep{Yahya:2014yva}, and Euclid \citep{EuclidTheoryWorkingGroup:2012gxx, Audren:2012vy,Harrison:2016stv}. The Euclid mission observes in the visible and near-infrared wavelength ranges.}
    \label{tab:instruspecs}
\end{table}

In real observations, several additional effects modify the theoretical power spectrum. These include redshift-space distortions (RSD) \citep{Kaiser:1987qv}, instrumental resolution effects and the Alcock–Paczynski (AP) effect \citep{Alcock:1979mp,Seo:2003pu}, arising from the assumed fiducial cosmology. Incorporating these contributions, the observed 21-cm power spectrum can be written as \citep{Sprenger:2018tdb},
\begin{equation}
    P_{21}(k,\mu,z) = f_{\rm AP}(z)\,f_{\rm res}(k,\mu,z)\,f_{\rm RSD}(k,\mu,z)\,b_{21}^{2}(z)\,P_{\rm m}(k,z),
\end{equation}
where $\mu$ denotes the cosine of the angle between the Fourier mode $k$ and the line-of-sight direction. Here, the redshift-space distortion term is given by $f_{\mathrm{RSD}}(k,\mu,z) = \left(1 + \beta(z)\mu^2\right)^2$, where $\beta(z) = f(z)/b(z)$ and $f(z) = d\ln D/d\ln a$ is the linear growth rate. The AP factor is $f_{\mathrm{AP}}(z) = \frac{D_{A,{\mathrm{fid}}}^2(z) H(z)}{{D_A}^2(z) H^{\mathrm{fid}}(z)}$, accounting for the mismatch between the fiducial and true cosmologies, where $H(z)$ is the Hubble parameter and $D_A(z)$ is the angular diameter distance. The factor $f_{\mathrm{res}}(k,\mu,z)$ accounts for instrumental resolution and residual systematics, including beam smoothing and redshift uncertainties. Observations would be dominated by telescope noise and foreground contamination \citep{Bonaldi:2025gpa}. Assuming the foregrounds to be spectrally smooth - and hence removable -  the telescope noise, $P_{\rm N}(z)$, contributes to the observed signal as follows,
\begin{equation}
    P_{21}^{\rm obs}(k,\mu,z)=P_{21}(k,\mu,z)+P_{\rm N}(z).
\end{equation}
For SKA intensity mapping surveys operating in single-dish mode, the instrumental noise power spectrum can be expressed as \citep{Villaescusa-Navarro:2016kbz,Spina:2021drb},
\begin{equation}
    P_{\rm N}(z)=T_{\rm sys}^{2}\frac{4\pi f_{\rm sky} r^{2}(z)(1+z)^{2}}{2H(z)t_{\rm tot}\nu_{0}N_{\rm dish}},
\end{equation}
where $T_{\rm sys}$ denotes the system temperature, $t_{\rm tot}$ is the total observing time, $N_{\rm dish}$ is the number of dishes, $r(z)$ is the comoving distance and $\nu_0$ is the rest frame frequency of the 21-cm hyperfine transition of neutral hydrogen. For the present work, $t_{\rm tot}$=1000 hours and $N_{\rm dish}$=200 were adopted as outlined in \citet{Villaescusa-Navarro:2016kbz}. Moreover, a reduced sky fraction of $f_{\rm sky}=0.58$ has been considered and a narrower frequency range \citep{Alonso:2014dhk} is accounted for to address foreground removal. The instrumental specifications for SKA-mid intensity mapping are listed in Table \ref{tab:instruspecs}.

\subsubsection{Galaxy Clustering}

Galaxy redshift surveys provide measurements of the three-dimensional distribution of galaxies, which act as biased tracers of the underlying matter density field. Similar to the 21-cm power spectrum in the post-reionization epoch, the observed galaxy power spectrum is related to the matter power spectrum through,
\begin{equation}
    P_{\rm g}(k,\mu,z)=f_{\rm AP}(z)\,f_{\rm res}(k,\mu,z)\,f_{\rm RSD}(k,\mu,z)\,b^{2}(z)\,P_{\rm m}(k,z),
\end{equation}
where $b(z)$ represents the galaxy bias. Both spectroscopic galaxy surveys from the Euclid mission as well as HI galaxy surveys expected from SKA-mid were considered for the present study.

Euclid will conduct a spectroscopic survey of approximately $10^{7}$ galaxies in the redshift range $0.7<z<2.0$ over a sky fraction $f_{\rm sky}=0.3636$. The number of detectable galaxies by Euclid for a redshift bin of width $\Delta z$ around $\Bar{z}$ is given as,
\begin{equation}
    N(\Bar{z}) = 41253 ~ f_{\rm sky} ~{\rm deg}^{2} \int_{\Bar{z}-{\rm \Delta} z/2}^{\Bar{z}+{\rm \Delta} z/2} \dfrac{{\rm dN}(z)/{\rm d}z}{1~\rm deg^{2}} {\rm d}z.
\end{equation}
The galaxy number count ${\rm dN}(z)/{\rm d}z$ is taken from table 3 of \citet{Pozzetti:2016cch} (``Model 1"), corresponding to a limiting flux of $3 \times 10^{-6} \ {\rm erg} \ {\rm s}^{-1} \ {\rm cm}^{-2}$. The expected redshift measurement uncertainty is taken to be $\sigma_{z}=0.001(1+z)$ \citep{EuclidTheoryWorkingGroup:2012gxx,Audren:2012vy}. Table \ref{tab:instruspecs} presents the detailed specifications for this mission. To account for uncertainties in galaxy bias modelling, nuisance parameters were introduced that modify the bias according to,
\begin{equation}
    b(z)=\beta_{0}^{\rm Euclid}(1+z)^{\beta_{1}^{\rm Euclid}},
\end{equation}
with Gaussian priors, $\mathcal{N}\sim(1,0.00625)$, applied to the nuisance parameters $\beta_{0}^{\rm Euclid}$ and $\beta_{1}^{\rm Euclid}$.

The most promising galaxy surveys for SKA are the SKA1-mid Band 1 (SKA1) and SKA1-mid Band 2 (SKA2). The survey volume was considered to be $S_{\rm area} = f_{\rm sky} \times 41253~ \rm deg^{2}$ according to the SKA baseline specifications \citep{Yahya:2014yva}. The observed galaxy number density and bias were obtained using fitting functions derived from simulations \citep{Yahya:2014yva}. The galaxy number count was modelled as,
\begin{equation}
    \dfrac{{\rm dN}(z)/{\rm d}z}{1~ \rm deg^{2}} = 10^{\rm c_{1}} z^{\rm c_{2}} \exp(-{\rm c_{3}}z),
    \label{equ.galaxy}
\end{equation}
while the HI galaxy bias is given by,
\begin{equation}
    b_{\rm HI}(z)={\rm c_{4}}\exp({\rm c_{5}}z).
    \label{equ.bias}
\end{equation}

The $c_i$ parameters are fitting parameters that are kept fixed in this analysis, with their values adopted from Table~6 of Ref.~\citep{Bull:2015lja}. Nuisance parameters modifying the bias were introduced in a manner similar to the Euclid case:
\begin{equation}
    b_{\rm HI}(z)=c_{4}\beta_{0}^{\rm SKA1/2}\exp\left(c_{5}\beta_{1}^{\rm SKA1/2}z\right),
\end{equation}
where Gaussian priors, $\mathcal{N}\sim(1,0.00625)$, were assigned to the nuisance parameters $\beta_{0}^{\rm SKA1/2}$ and $\beta_{1}^{\rm SKA1/2}$. Further specifications for SKA1 and SKA2 are outlined in Table \ref{tab:instruspecs}. 

\subsubsection{Cosmic Shear}

Weak gravitational lensing provides a direct probe of the projected matter distribution through distortions in the shapes of background galaxies. The tomographic cosmic shear power spectrum between redshift bins $i$ and $j$ for multipole $l$ is given by \citep{Sprenger:2018tdb},
\begin{equation}
    C_{l}^{ij}=\frac{9}{16}\Omega_{m}^{2}H_{0}^{4}\int_{0}^{\infty}\frac{dr}{r^{2}}g_{i}(r)g_{j}(r)P\!\left(\frac{l}{r},z(r)\right),
\end{equation}
where $P\left(k=\dfrac{l}{r},z(r)\right)$ is the three dimensional matter power spectrum, and $g_{i}(r)$ denote the convolution of the distribution of the observed galaxies with redshift error \citep{Dey:2023sxx}. Moreover, the random intrinsic alignment of the galaxies introduces an additional noise contribution \citep{Hu:2003pt},
\begin{equation}
    N_{l}^{ij}=\delta_{ij}\frac{\sigma_{g_{i}}^{2}}{n_{i}},
\end{equation}
where $\sigma_{g_{i}}$ is the root-mean-square intrinsic ellipticity and $n_{i}$ is the galaxy number density in the $i$th redshift bin \citep{Harrison:2016stv}. In this analysis, cosmic shear observations from both Euclid and SKA have been considered. The specifications of Euclid and SKA are summarized in Table \ref{tab:instruspecs}.

\subsubsection{Conservative and Realistic Prescriptions}

To account for theoretical uncertainties arising from non-linear structure formation, scale cuts were imposed on the Fourier modes included in the analysis. On large scales a minimum wavenumber $k_{\rm min}=0.02\,{\rm Mpc}^{-1}$ was taken, corresponding to the largest scale accessible given the width of the redshift bins, below which the small-angle approximation breaks down.

On small scales two different prescriptions were adopted. The first is a conservative prescription \citep{Sprenger:2018tdb}, in which the analysis is restricted to scales below the redshift-dependent non-linear scale,
\begin{equation}
    k_{\rm NL}(z)=k_{\rm NL}(0)(1+z)^{\frac{2}{2+n_{s}}}.
\end{equation}

For SKA galaxy clustering and 21-cm intensity mapping observations, $k_{\rm NL}(0)=0.2\,h\,{\rm Mpc}^{-1}$ was adopted. For SKA cosmic shear measurements multipole cuts were applied between $l_{\rm min}=5$ and a redshift-dependent maximum multipole given by $l_{\rm max}^{i}=k_{\rm NL}(z)\bar{r}_{\rm peak}^{i}$, where $k_{\rm NL}(0)=0.5\,h\,{\rm Mpc}^{-1}$.

The second prescription adopted a more optimistic or realistic treatment of non-linear scales \citep{Sprenger:2018tdb}. In this case the analysis was extended up to $k_{\rm max}=10\,h\,{\rm Mpc}^{-1}$ for SKA galaxy clustering and 21-cm intensity mapping observations. For Euclid galaxy clustering, $k_{\rm max}$ was also taken to be $10\,h\,{\rm Mpc}^{-1}$. For SKA cosmic shear measurements $k_{\rm NL}(0)=2.0\,h\,{\rm Mpc}^{-1}$ was adopted, while for Euclid cosmic shear measurements, $k_{\rm NL}(0)=5.0\,h\,{\rm Mpc}^{-1}$ was adopted. Forecasts were performed for different combinations of these prescriptions.


\section{Results and Discussion}\label{sec:results}

\begin{table*}
    \resizebox{1.0\textwidth}{!}{\renewcommand{\arraystretch}{1.5} \setlength{\tabcolsep}{10 pt}
    \begin{tabular}{c c c c c c }
        \hline
        \textbf{Parameter} & \makecell{\textbf{CBS + SKA1 IM Band 1}\\(cons.)\\(real.)} & \makecell{\textbf{CBS + SKA1 IM Band 2}\\(cons.)\\(real.)} & \makecell{\textbf{CBS + SKA1 GC+CS}\\(cons.)\\(real.)} &
        \makecell{\textbf{CBS + SKA2 GC+CS}\\(cons.)\\(real.)} & \makecell{\textbf{CBS + Euclid GC+CS}}\\
        \hline
        \multirow{2}{*}{\boldmath${100\Omega_b}{h^2}$} & $2.251\pm 0.011            $ & $2.252\pm 0.012            $ & $2.250\pm 0.012            $ & $2.248\pm 0.011            $ & \multirow{2}{*}{$2.248\pm 0.011            $}\\
        & $2.250\pm 0.011            $ & $2.246\pm 0.012            $ & $2.250\pm 0.012            $ & $2.248\pm 0.011            $\\
        \hdashline
        \multirow{2}{*}{\boldmath${\Omega_c}{h^2}$} & $0.16372\pm 0.00079        $ & $0.16331\pm 0.00070        $ & $0.1626\pm 0.0015          $ & $0.16382\pm 0.00042        $ & \multirow{2}{*}{$0.16304\pm 0.00055        $}\\
        & $0.16404\pm 0.00078        $ & $0.16365\pm 0.00053        $ & $0.16340\pm 0.00088        $ & $0.16365\pm 0.00034        $\\
        \hdashline
        \multirow{2}{*}{\boldmath$100{\theta_s}$} & $1.04203\pm 0.00026        $ & $1.04205\pm 0.00027        $ & $1.04202\pm 0.00027        $ & $1.04195\pm 0.00024        $ & \multirow{2}{*}{$1.04192\pm 0.00028        $}\\
        & $1.04200\pm 0.00026        $ & $1.04199\pm 0.00027        $ & $1.04201\pm 0.00028        $ & $1.04194\pm 0.00025        $\\
        \hdashline
        \multirow{2}{*}{\boldmath${\ln{\left({10^{10}A_s}\right)}}$} & $3.038\pm 0.011            $ & $3.044\pm 0.014            $ & $3.044\pm 0.013            $ & $3.0333\pm 0.0059          $ & \multirow{2}{*}{$3.0313\pm 0.0067          $}\\
        & $3.040\pm 0.011            $ & $3.041\pm 0.011            $ & $3.043\pm 0.013            $ & $3.0300\pm 0.0040          $\\
        \hdashline
        \multirow{2}{*}{\boldmath$n_s$          } & $0.9695\pm 0.0025          $ & $0.9692\pm 0.0032          $ & $0.9687\pm 0.0033          $ & $0.9683\pm 0.0016          $ & \multirow{2}{*}{$0.96663\pm 0.00072        $}\\
        & $0.9686\pm 0.0026          $ & $0.9676\pm 0.0028          $ & $0.9686\pm 0.0022          $ & $0.96826\pm 0.00050        $\\
        \hdashline
        \multirow{2}{*}{\boldmath${\tau}$       } & $0.0529\pm 0.0057          $ & $0.0554\pm 0.0072          $ & $0.0555\pm 0.0071          $ & $0.0505\pm 0.0038          $ & \multirow{2}{*}{$0.0498\pm 0.0040          $}\\
        & $0.0540\pm 0.0058          $ & $0.0537\pm 0.0057          $ & $0.0552\pm 0.0067          $ & $0.0489\pm 0.0032          $\\
        \hdashline
        \multirow{2}{*}{\boldmath$Q$            } & $0.4820\pm 0.0089          $ & $0.4791\pm 0.0071          $ & $0.470\pm 0.020            $ & $0.4839\pm 0.0038          $ & \multirow{2}{*}{$0.4702\pm 0.0058          $}\\
        & $0.4868\pm 0.0094          $ & $0.4815\pm 0.0050          $ & $0.480\pm 0.011            $ & $0.4810\pm 0.0023          $\\
        \hdashline
        \multirow{2}{*}{\boldmath$w_0$          } & $-1.019^{+0.013}_{-0.0087} $ & $-1.0169\pm 0.0044         $ & $-1.031^{+0.029}_{-0.0085} $ & $-1.0170\pm 0.0039         $ & \multirow{2}{*}{$> -1.02                   $}\\
        & $-1.020^{+0.011}_{-0.0086} $ & $-1.0143\pm 0.0024         $ & $-1.021^{+0.018}_{-0.0079} $ & $-1.0155\pm 0.0028         $\\
        \hdashline
        \multirow{2}{*}{\boldmath$w_a$          } & $-0.462\pm 0.032           $ & $-0.437\pm 0.034           $ & $-0.408^{+0.079}_{-0.088}  $ & $-0.466\pm 0.019           $ & \multirow{2}{*}{$-0.513\pm 0.038           $}\\
        & $-0.458\pm 0.031           $ & $-0.474\pm 0.018           $ & $-0.430\pm 0.057           $ & $-0.478\pm 0.012           $\\
        \hline
        \multirow{2}{*}{\boldmath$H_0$          } & $67.94\pm 0.15             $ & $67.73\pm 0.12             $ & $67.83\pm 0.31             $ & $67.663\pm 0.082           $ & \multirow{2}{*}{$68.22\pm 0.18             $}\\
        & $67.83\pm 0.16             $ & $67.612\pm 0.092           $ & $67.66\pm 0.16             $ & $67.736\pm 0.073           $\\
        \hdashline
        \multirow{2}{*}{\boldmath$\Omega_{m0}$  } & $0.4049\pm 0.0029          $ & $0.4065\pm 0.0024          $ & $0.4038\pm 0.0061          $ & $0.4083\pm 0.0012          $ & \multirow{2}{*}{$0.4000\pm 0.0020          $}\\
        & $0.4069\pm 0.0030          $ & $0.4085\pm 0.0017          $ & $0.4076\pm 0.0032          $ & $0.40707\pm 0.00073        $\\
        \hdashline
        \multirow{2}{*}{\boldmath$\sigma_{8,0}$ } & $0.6283\pm 0.0018          $ & $0.6292\pm 0.0033          $ & $0.6346\pm 0.0077          $ & $0.6288\pm 0.0014          $ & \multirow{2}{*}{$0.6319\pm 0.0018          $}\\
        & $0.6280\pm 0.0019          $ & $0.6325\pm 0.0022          $ & $0.6306\pm 0.0034          $ & $0.62888\pm 0.00076        $\\
        \hdashline
        \multirow{2}{*}{\boldmath$S_8$          } & $0.7299\pm 0.0024          $ & $0.7325\pm 0.0040          $ & $0.7361\pm 0.0048          $ & $0.73362\pm 0.00077        $ & \multirow{2}{*}{$0.7296\pm 0.0017          $}\\
        & $0.7313\pm 0.0023          $ & $0.7381\pm 0.0023          $ & $0.7350\pm 0.0022          $ & $0.73256\pm 0.00040        $\\
        \hline
    \end{tabular}
    }
    \caption{Projected constraints from the combined analysis of SKA intensity mapping, galaxy clustering and cosmic shear observations, with current CMB+BAO+SNIa data, for the i-CPL phantom model. Results are shown for SKA1 and SKA2 surveys under both conservative and realistic treatments of non-linear scales, and for Euclid.}
    \label{tab:main_constraints}
\end{table*}

\begin{figure}
    \centering
    \includegraphics[width=0.49\textwidth]{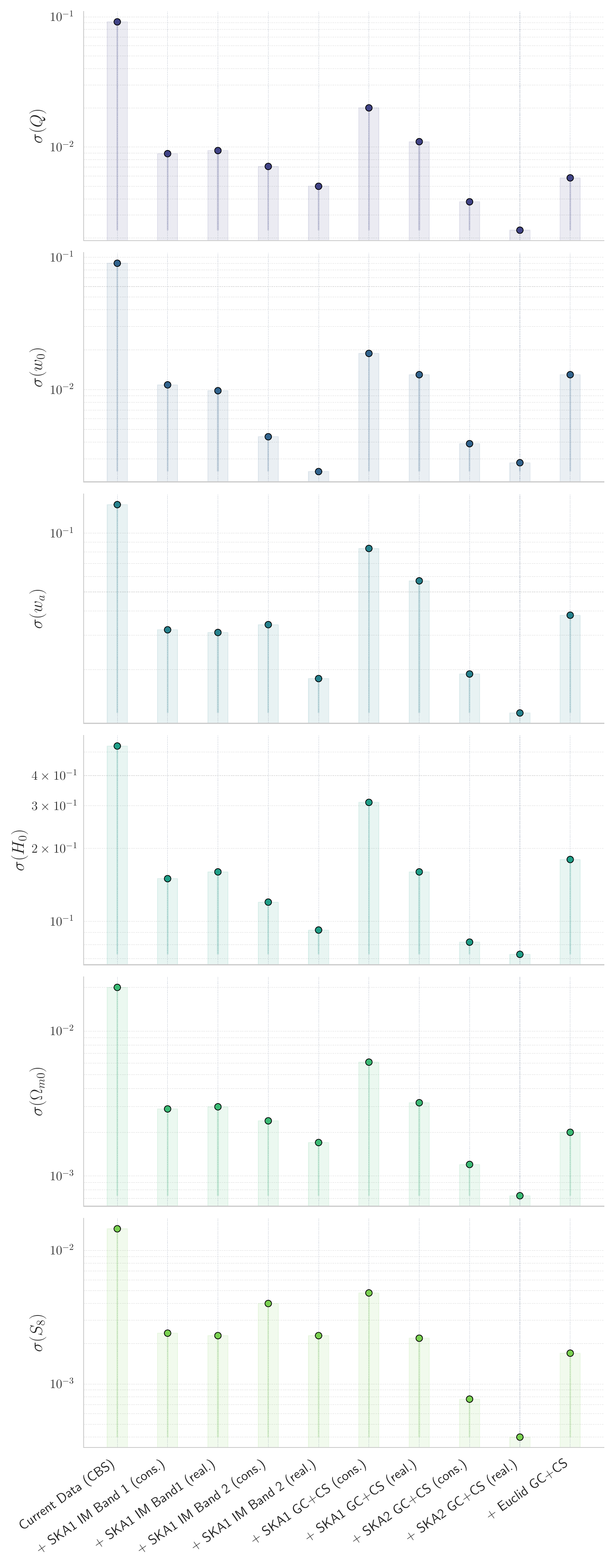}
    \caption{Projected $1\sigma$ errors for SKA intensity mapping, galaxy clustering and cosmic shear observations for the i-CPL phantom model. Results are shown for SKA1 and SKA2 surveys under both conservative and realistic treatments of non-linear scales, and for Euclid.}
    \label{fig:errors}
\end{figure}

\begin{figure*}
    \centering
    \includegraphics[width=0.49\textwidth]{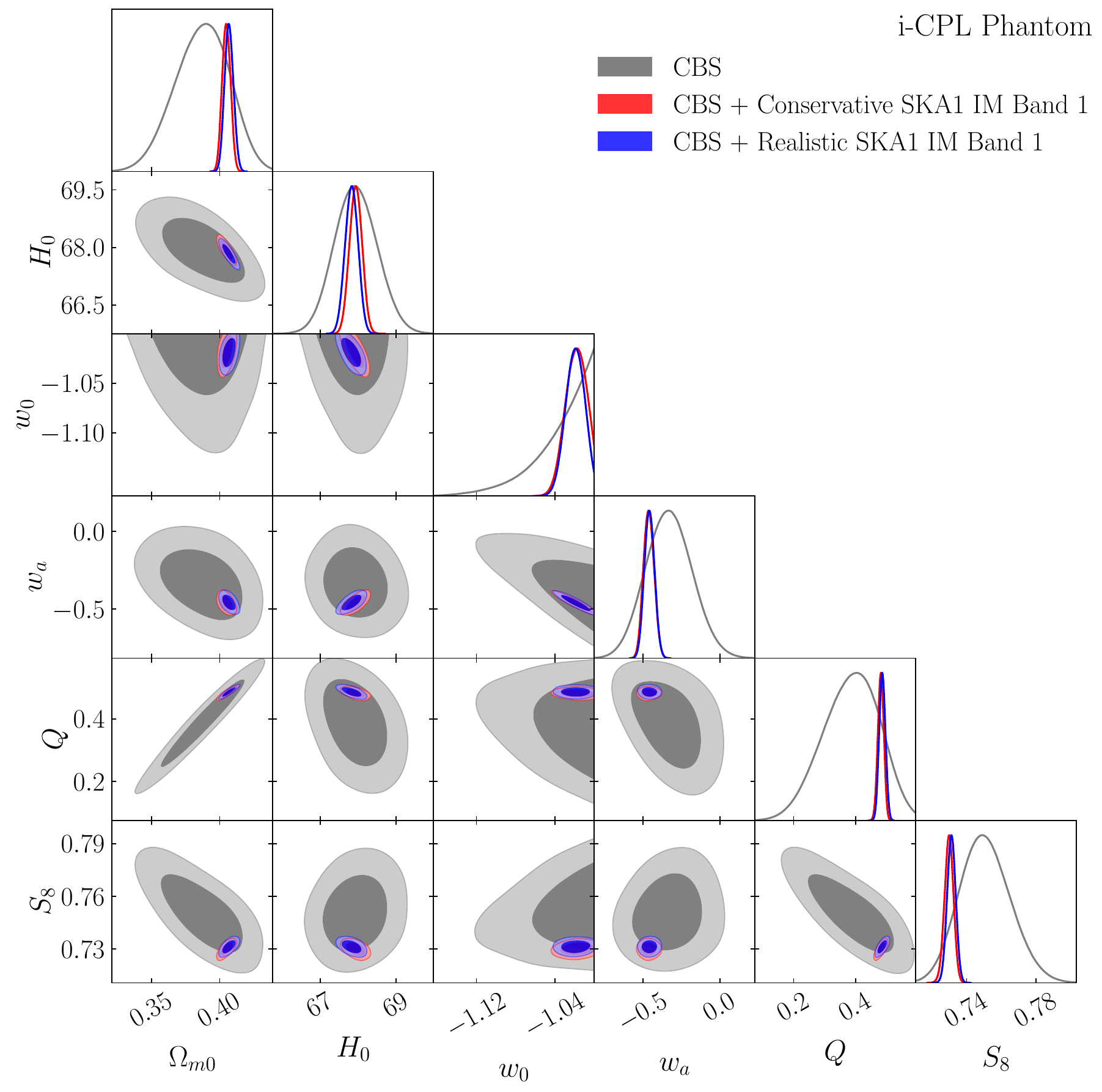}
    \includegraphics[width=0.49\textwidth]{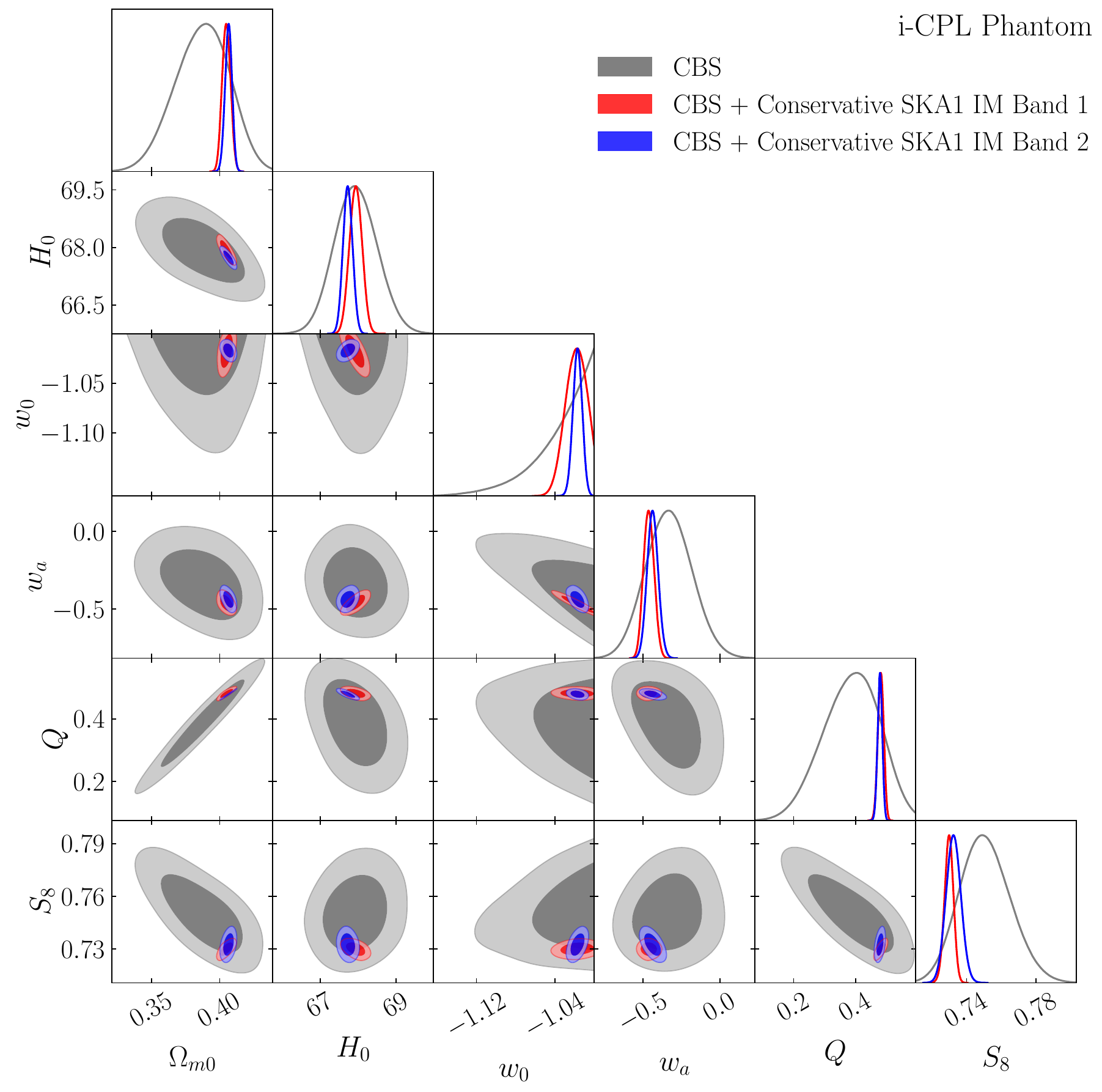}
    \caption{Projected constraints for SKA1 intensity mapping observations for the i-CPL phantom model. The figure compares the constraints from current CMB+BAO+SNIa data with (left panel) the projected SKA1 Band 1 forecasts under the conservative and realistic treatments of non-linear scales, and (right panel) from SKA1 intensity mapping observations in Band~1 and Band~2 under the conservative prescription.}
    \label{fig:IM}
\end{figure*}

\begin{figure*}
    \centering
    \includegraphics[width=0.49\textwidth]{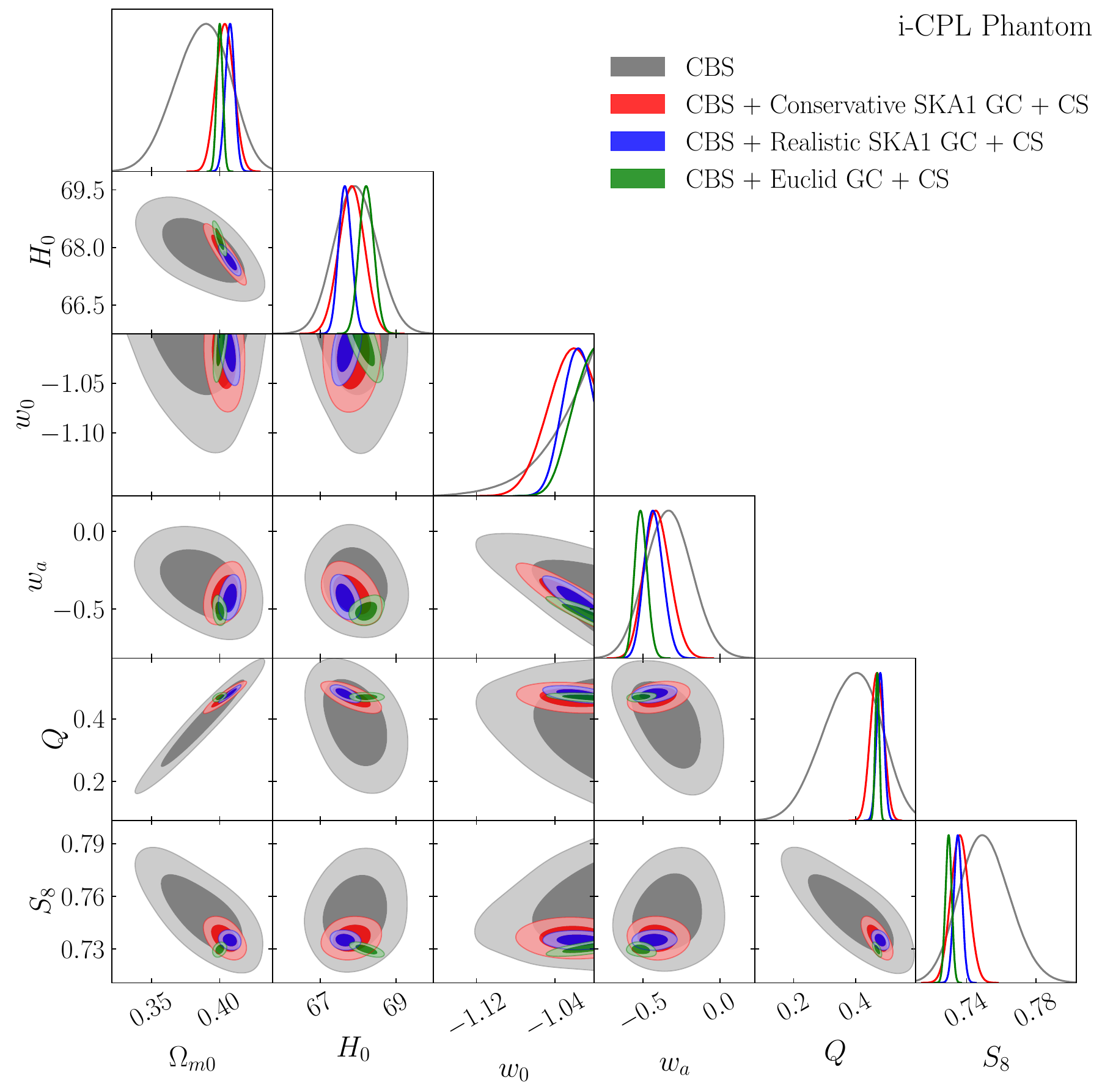}
    \includegraphics[width=0.49\textwidth]{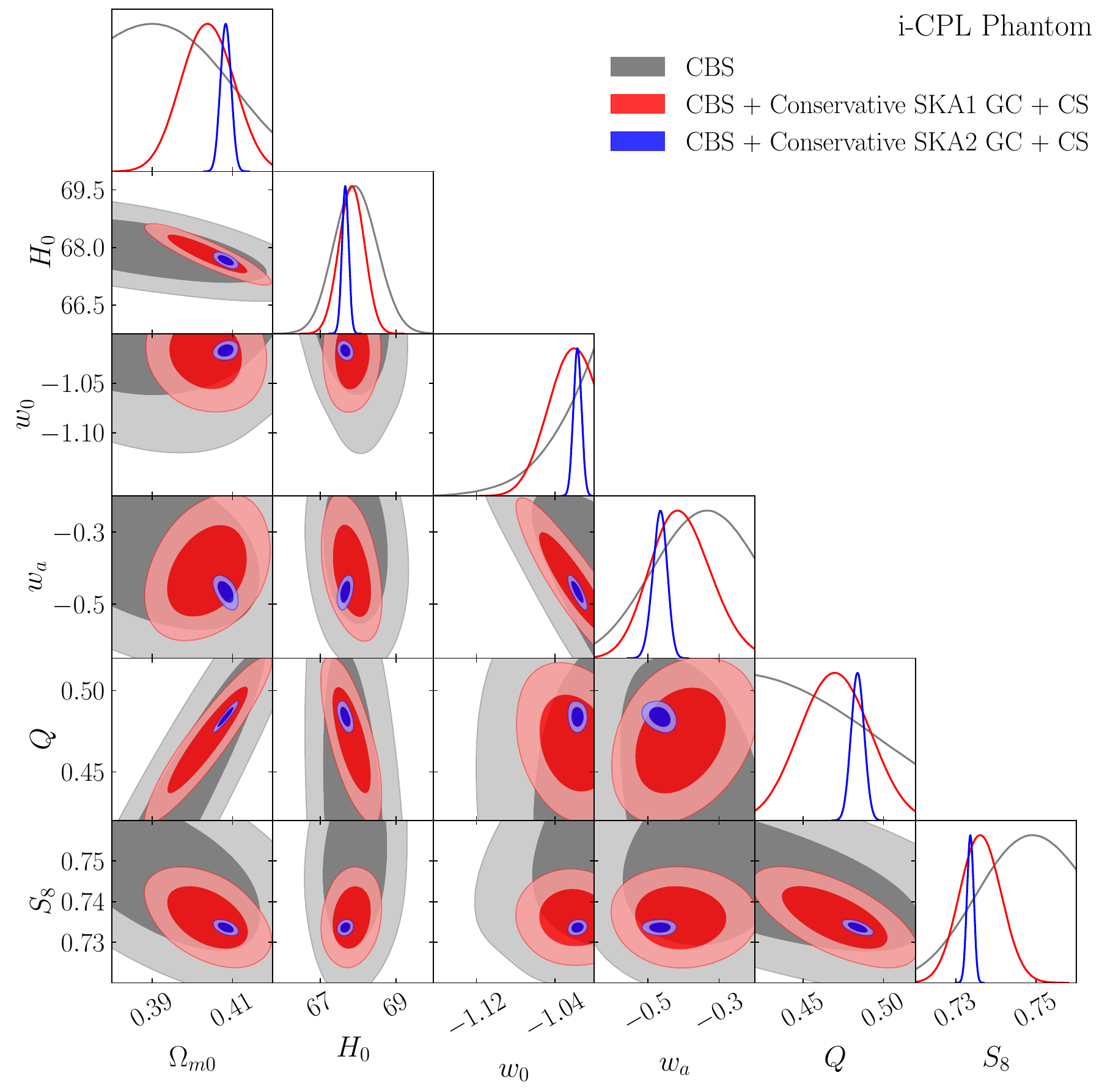}
    \caption{Projected constraints for SKA-mid galaxy clustering and cosmic shear observations for the i-CPL phantom model. The figure compares the constraints from current CMB+BAO+SNIa data with (left panel) the projected SKA1 forecasts, under the conservative and realistic treatments of non-linear scales, and Euclid, and (right panel) from SKA1 and SKA2 under the conservative prescription.}
    \label{fig:SKA_combined}
\end{figure*}

As discussed earlier, the interacting scenario is first explored with the help of the DE EoS parametrized as $w(a) = w_0 + w_a (1-a)$ \citep{Chevallier:2000qy,Linder:2002et}, commonly known as the CPL parametrization. Further, the EoS is restricted to the phantom regime ($w<-1$), since earlier studies have shown this scenario to be helpful in the context of the $S_8$ tension \citep{Shah:2024rme}. This model serves as the primary fiducial framework for the forecasts.

The baseline constraints for this model were obtained from current observations, reproduced from \citet{Shah:2025ayl} in Table \ref{tab:current} of Appendix \ref{appendix}. These results provide the best-fitting cosmological parameter values (Table \ref{tab:bestfit}) used to generate the mock datasets employed in the present analysis.

First, the constraining power of several probes: 21-cm intensity mapping with SKA-mid and galaxy clustering and cosmic shear from SKA-mid and Euclid, is examined. Throughout this section, the projected constraints from these future surveys are compared with the baseline constraints obtained from current CMB+BAO+SNIa data to quantify the expected improvement in parameter uncertainties. Fig. \ref{fig:errors} compares the errors on the parameters $Q$, $w_0$, $w_a$ $H_0$, $\Omega_m$ and $S_8$ for the different data combinations. The forecasted constraints for all parameters are listed in Table \ref{tab:main_constraints}, while the corresponding contour plots are in Fig. \ref{fig:IM} and Fig. \ref{fig:SKA_combined}. The forecasted constraints are consistent with the fiducial inputs to within $2\sigma$. This indicates that the generated mock datasets are consistent with the underlying fiducial cosmology derived from current observations. The main features of the forecasted constraints can be summarized as follows:

\begin{itemize}
    \item \textbf{Nature of the Interaction:} As previously shown by \citet{Shah:2024rme}, a phantom DE EoS tends to favour a positive value of the interaction parameter, driven by the negative correlation between $Q$ and $w_0$. The forecasts are consistent with this trend and suggest that future surveys could significantly improve the precision on $Q$. In particular, within the forecasting framework presented here, SKA1 IM Band 1 observations indicate potential improvements by a factor of $\sim 2$ in the conservative case and up to $\sim 10$ in the realistic case. Comparable trends are seen for Band 2, with the most optimistic configuration reaching factors of $\sim 20$. The SKA1 GC+CS combination is forecasted to improve constraints by factors of $\sim 5$–$10$, while SKA2 GC+CS could yield substantially tighter constraints, approaching $\sim 40$ in the most optimistic scenario. Euclid GC+CS is expected to provide an improvement at the level of a factor of $\sim 15$.
    \item \textbf{Evolution of Dark Energy:} The DE EoS parameters $w_0$ (present day value) and $w_a$ (evolution parameter) are forecasted to be reasonably tightly constrained by future SKA-mid observations. SKA1 IM Band 1 and Band 2 could improve constraints by factors of $\sim 2$ and $\sim 10$ in the conservative case, with realistic configurations providing a further improvement by a factor of $\sim 2$. The SKA1 GC+CS combination is expected to yield improvements at the level of $\sim 1.5$–$2.5$, while SKA2 GC+CS could further tighten these constraints, reaching up to $\sim 11$ in the most optimistic case. Euclid GC+CS is likewise forecast to improve constraints on $w_0$ and $w_a$ by factors of $\sim 2.5$–$3.5$
    \item \textbf{Constraints on Background Parameters:} Substantial improvements in background cosmological parameters are forecasted, in some cases approaching one to two orders of magnitude, particularly for the realistic SKA2 GC+CS configuration. More generally, realistic prescriptions tend to yield approximately a factor of $\sim 2$ tighter constraints than conservative ones. Within SKA1, intensity mapping provides somewhat tighter constraints than the GC+CS combination, whereas SKA2 GC+CS forecast yields constraints comparable to, or better than, those of SKA1 IM. The increase in precision is dominantly driven by tighter constraints on $\Omega_c h^2$, while $\Omega_b h^2$ demonstrates comparatively limited improvement. This is because the late-time matter power spectrum is especially sensitive to the total matter density through its broadband shape. Although baryons also affect both the transfer function and the BAO features, the dominant contribution to the total matter density comes from cold dark matter, leading to tighter constraints on $\Omega_ch^2$. Since $\Omega_bh^2$ is already tightly constrained by CMB observations, with additional independent information from big-bang nucleosynthesis (BBN), the incremental improvement from late-time LSS data is comparatively modest. The precision of $\theta_s$ (primarily set by pre-recombination acoustic physics) remains largely unchanged, as it is the baryonic imprint on the surface of last scattering, and is weakly affected by late-time changes in clustering. The Hubble constant $H_0$ is forecasted to be constrained significantly, with SKA1 IM potentially reducing uncertainties by factors of $\sim 2$–$5$, and SKA2 GC+CS reaching improvements of up to $\sim 70$ under optimistic assumptions. This improvement is primarily driven my tightening of constraints on $\Omega_c h^2$, which is correlated with $H_0$, and also due to higher precision in the DE EoS parameters. As for Euclid GC+CS, the forecasted constraints are comparable to the SKA1 GC+CS (realistic) case.
    \item \textbf{Clustering Parameters:} The clustering parameters $\sigma_{8,0}$ and $S_8$ are forecast to be particularly well constrained by LSS observations. All survey configurations lead to at least an order of magnitude reduction in uncertainties, with realistic prescriptions typically improving upon conservative ones by roughly a factor of $\sim 2$. The SKA2 GC+CS configuration yields the strongest projected improvement, reaching up to $\sim 30$ in the most optimistic case.
    \item \textbf{Primordial Parameters:} The primordial parameters $A_s$ and $n_s$ mostly demonstrate comparatively modest improvements in the forecasts. Only the SKA2 GC+CS (realistic) configuration and Euclid are forecasted to improve precision by factors of $\sim 3.5$ and $\sim 7$, respectively. $A_s$, contributes to the overall normalization of the matter power spectrum and can be constrained through LSS observations, although not as strongly as by CMB measurements. In contrast, the spectral index $n_s$ has limited impact on the late-time observables considered here and are therefore not expected to yield significant improvement from these surveys. The improvements mainly arise from the overall tightening of constraints on other cosmological parameters.
\end{itemize}

It is noteworthy that SKA1 IM Band 1 provides relatively weaker constraints in general compared to Band 2, despite probing a larger survey volume (see Table \ref{tab:instruspecs}). This can be understood from the structure of the interacting dark sector model considered here, in which the interaction term scales as $Q\propto\rho_{de}$. Since dark energy becomes dominant only at low redshifts, the impact of the interaction and of the DE EoS is significantly enhanced in this regime. As a result, observations targeting lower redshifts, such as those from Band 2, are more sensitive to these parameters and therefore yield tighter constraints.

Furthermore, realistic configurations yield significantly tighter constraints than conservative ones, primarily by incorporating a broader range of scales. Including smaller-scale (high-$k$) modes - typically extending up to $k_{\rm max} \approx 10 \, h \, \text{Mpc}^{-1}$ compared to a conservative limit of $k_{\rm max} \approx 0.2 \, h \, \text{Mpc}^{-1}$ \citep{Sprenger:2018tdb,SKA:2018ckk}, especially in the case of SKA intensity mapping and galaxy clustering observations - substantially increases the number of independent Fourier modes, thereby boosting statistical power. These scales are particularly sensitive to the growth of structure and matter perturbations, which are key indicators of the dark matter dark energy interaction strength. However, their use is limited by theoretical uncertainties: at higher wavenumbers ($k \gtrsim 0.15 \, h \, \text{Mpc}^{-1}$), non-linear gravitational collapse, baryonic feedback and scale-dependent bias become significant \citep{Santos:2015gra}. While the conservative approach ensures theoretical robustness by restricting the analysis to the quasi-linear regime, it discards a significant fraction of available information, resulting in weaker constraints. In contrast, the realistic configuration includes mildly non-linear scales, assuming that future improvements in modelling and systematic control will enable their reliable use. Together, these two cases effectively bracket the expected performance of upcoming surveys.

\subsection{The Non-Phantom Regime}

\begin{table}
    \resizebox{0.5\textwidth}{!}{\renewcommand{\arraystretch}{1.2} \setlength{\tabcolsep}{40 pt}
    \begin{tabular}{c c}
        \hline
        \textbf{Parameter} & \textbf{CBS + SKA1 GC+CS} \\
        \hline
        {\boldmath${100\Omega_b}{h^2}$} & $2.252\pm 0.012            $\\
        {\boldmath${\Omega_c}{h^2}$} & $0.03719^{+0.00040}_{-0.00036}$\\
        {\boldmath$100{\theta_s}$} & $1.04208\pm 0.00027        $\\
        {\boldmath${\ln{\left({10^{10}A_s}\right)}}$} & $3.0565\pm 0.0083          $\\
        {\boldmath$n_s$          } & $0.9712\pm 0.0023          $\\
        {\boldmath${\tau}$       } & $0.0615\pm 0.0050          $\\
        {\boldmath$Q$            } & $-0.6030\pm 0.0063         $\\
        {\boldmath$w_0$          } & $-0.7259\pm 0.0086         $\\
        {\boldmath$w_a$          } & $-0.239^{+0.029}_{-0.039}  $\\
        \hline
        {\boldmath$H_0$          } & $67.65\pm 0.18             $\\
        {\boldmath$\Omega_{m0}$  } & $0.1319\pm 0.0010          $\\
        {\boldmath$\sigma_{8,0}$ } & $1.7888\pm 0.0054          $\\
        {\boldmath$S_8$          } & $1.1860\pm 0.0040          $\\
        \hline
    \end{tabular}
    }
    \caption{Projected constraints for SKA1 galaxy clustering and cosmic shear observations for the i-CPL non-phantom model ($w>-1$), with realistic treatment of non-linear scales.}
    \label{tab:CPL_quint_constraints}
\end{table}

\begin{figure}
    \centering
    \includegraphics[width=0.49\textwidth]{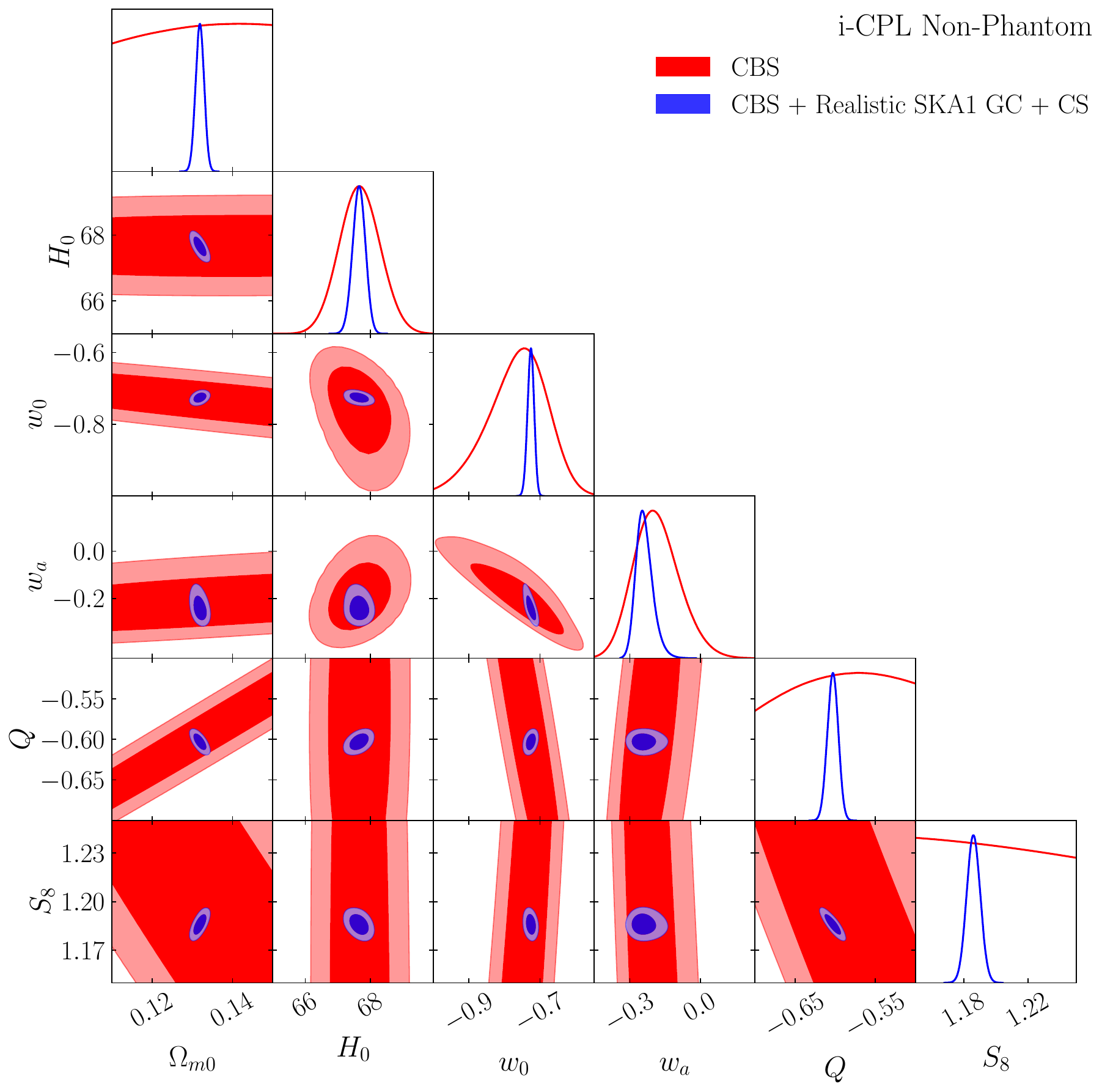}
    \caption{Projected constraints for SKA1 galaxy clustering and cosmic shear observations for the i-CPL non-phantom model ($w>-1$). The figure compares the constraints from current CMB+BAO+SNIa data with the projected SKA1 forecasts under the realistic treatment of non-linear scales.}
    \label{fig:CPL_quint}
\end{figure}

The preceding analysis adopted the CPL parametrization in the phantom regime as the fiducial iDMDE model. This choice provided a useful benchmark for testing the relative constraining power of different future survey combinations. However, if the DE EoS is non-phantom from the outset, the sensitivity of projected constraints from future surveys to this change in the underlying model assumption must be assessed.

To investigate this possibility, the analysis was repeated for the CPL parametrization in the non-phantom regime, where the DE EoS satisfies $w>-1$ throughout cosmic evolution. Table \ref{tab:CPL_quint_constraints} and Fig. \ref{fig:CPL_quint} show the projected parameter constraints obtained from the combined SKA1 galaxy clustering and cosmic shear observations (with a realistic non-linear cut-off) in the non-phantom regime. These results are compared with the corresponding constraints derived from current observations (Planck 2018 + DESI DR2 BAO + Pantheon+).

We note that the extreme values of $\Omega_m$ and $\sigma_{8,0}$ arise due to strong parameter degeneracies involving the interaction strength $Q$ and the dark energy equation of state parameters. In particular, $w_0$ and $Q$ are negatively correlated, such that in the non-phantom regime, higher values of $w_0$ favour negative values of $Q$. From the background continuity equations, negative values of $Q$ correspond to energy transfer from dark matter to dark energy, causing the dark matter density to dilute faster than the standard $a^{-3}$ scaling and thereby favouring lower values of $\Omega_m$. At the perturbation level, the interaction terms modify the effective evolution of dark matter fluctuations, such that for negative values of $Q$, the effective friction term in the dark matter velocity perturbation equation is reduced, allowing matter perturbations to grow more efficiently and thereby leading to larger values of $\sigma_{8,0}$. Consequently, these correlated effects produce statistically allowed but non-standard parameter combinations, which makes the non-phantom scenario not as favoured as the phantom regime \citep{Shah:2024rme,Shah:2025ayl}.

The constraints indicate that SKA-mid will be capable of constraining iDMDE models irrespective of whether the DE EoS lies in the phantom or non-phantom regime. Although the allowed parameter regions differ slightly between the two cases due to the different parameter bounds, the overall improvement in parameter constraints from SKA-mid observations remains comparable in both scenarios.

\subsection{Effect of the Dark Energy Sound Speed}

\begin{figure}
    \centering
    \includegraphics[width=0.49\textwidth]{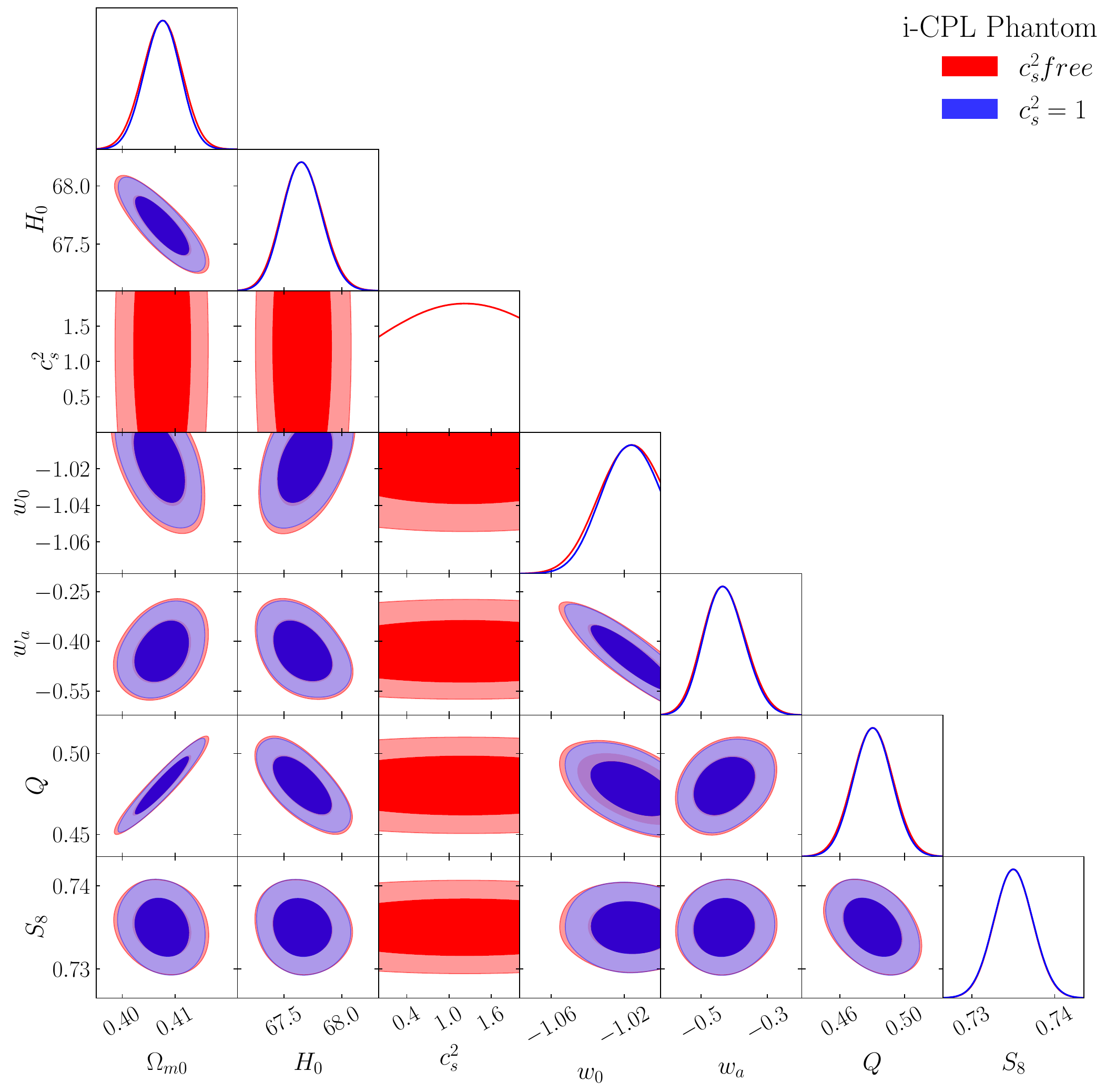}
    \caption{Projected constraints for SKA1 galaxy clustering and cosmic shear observations for the i-CPL phantom model with the sound speed of dark energy perturbations $c_s^2$ free, compared with $c_s^2=1$, under the realistic treatment of non-linear scales.}
    \label{fig:cs2_test}
\end{figure}

\begin{figure*}
    \centering
    \includegraphics[width=0.49\textwidth]{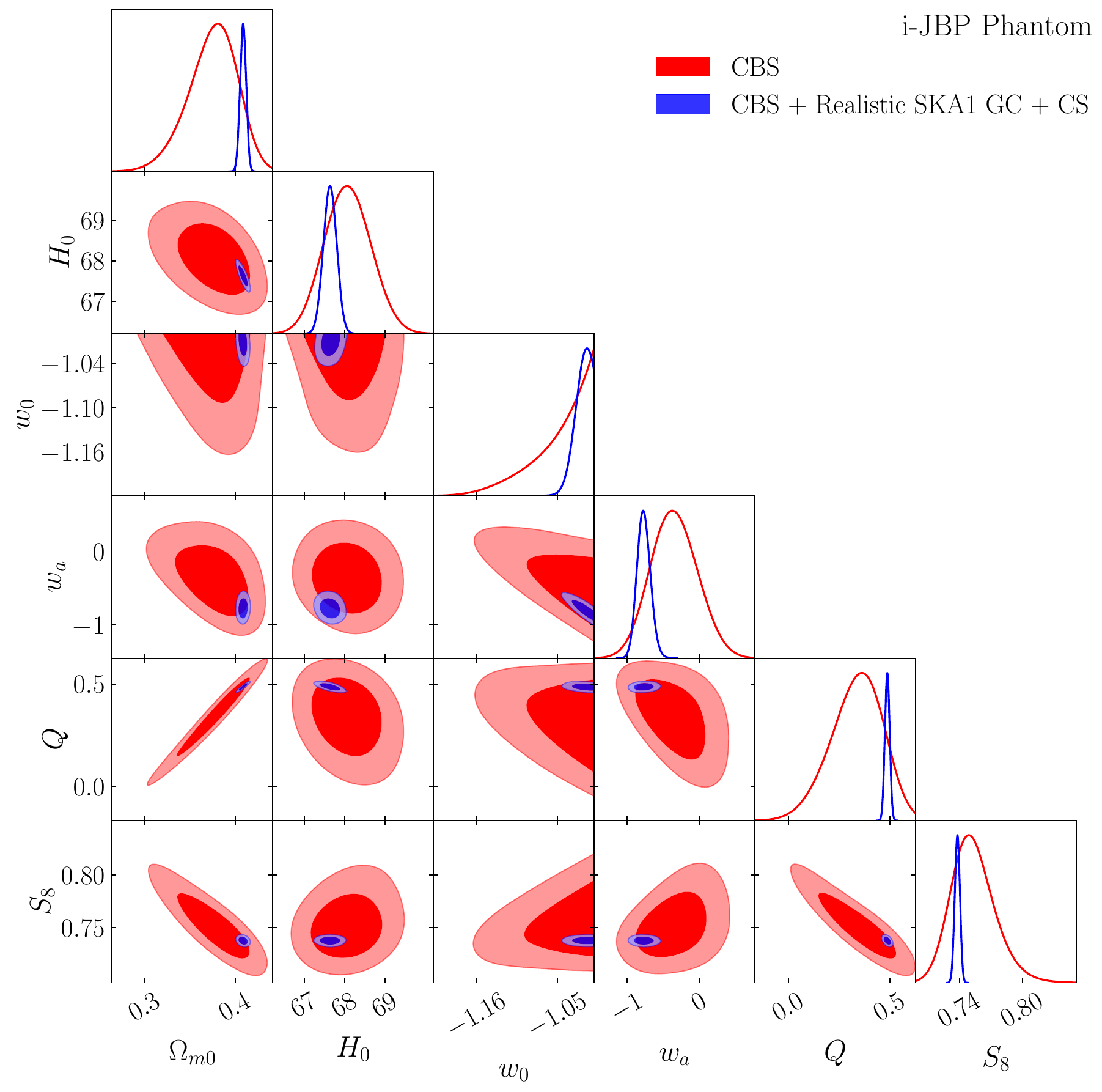}
    \includegraphics[width=0.49\textwidth]{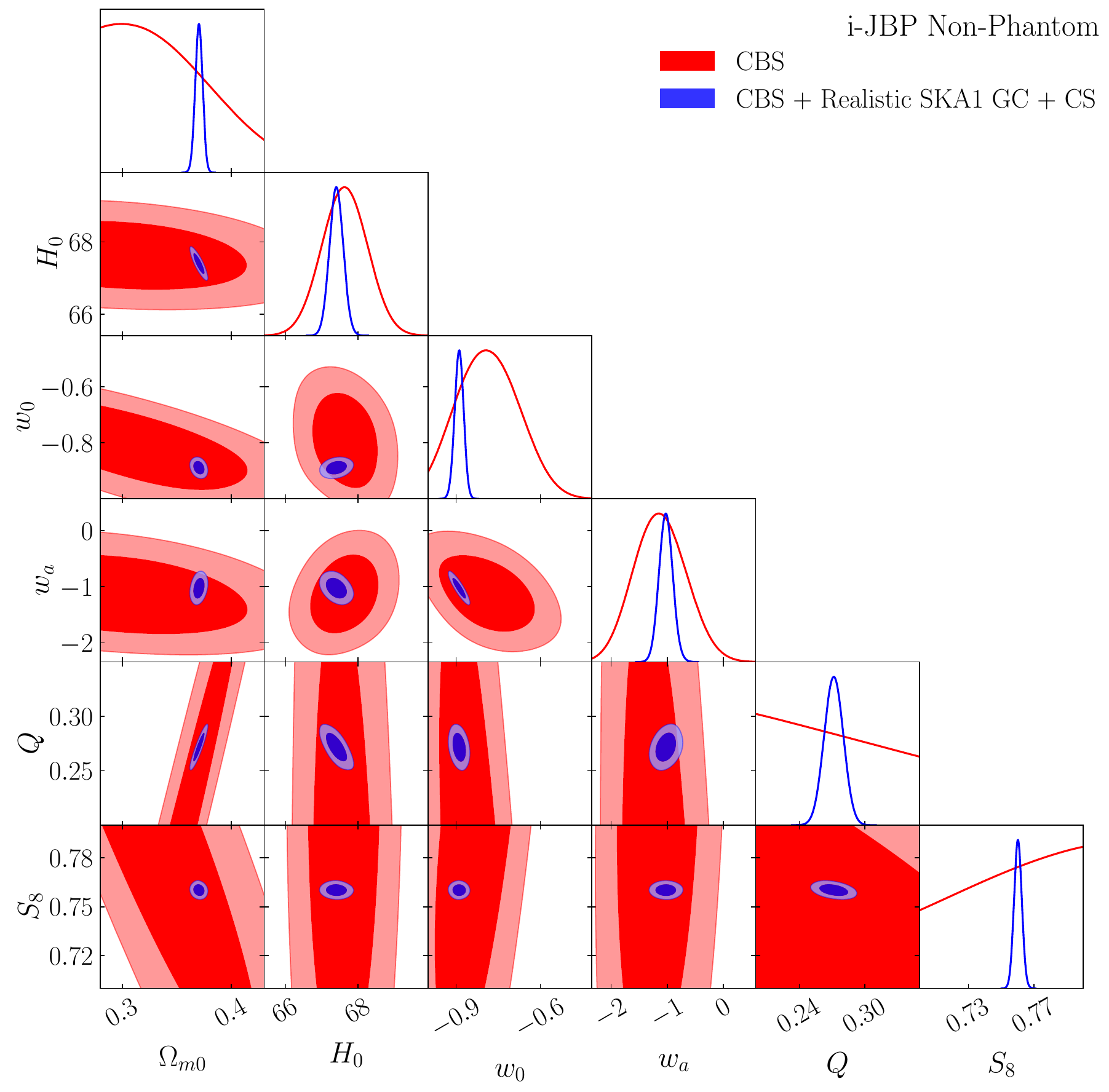}
    \caption{Projected constraints for SKA1 galaxy clustering and cosmic shear observations for the i-JBP phantom and non-phantom models. The figure compares the constraints from current CMB+BAO+SNIa data with the projected SKA1 forecasts under the realistic treatment of non-linear scales.}
    \label{fig:JBP}
\end{figure*}

In the analyses presented above, the sound speed of dark energy perturbations was fixed to $c_s^2 = 1$. However, the clustering properties of dark energy can, in principle, depend on the value of this parameter, which motivates testing the robustness of the results against this assumption.

To examine this effect, the analysis for the CPL phantom scenario was repeated while allowing the dark energy sound speed $c_s^2$ to vary as a free parameter. Fig. \ref{fig:cs2_test} shows the projected parameter constraints obtained from the combined SKA1 galaxy clustering and cosmic shear observations with a realistic non-linear cut-off.  

Although $c_s^2=1$ was used for generating the mock data, no meaningful constraints on $c_s^2$ were found, with the posterior distribution remaining largely unconstrained across the prior range considered ($c^2_{s}\in[0,2]$). This behaviour is expected, since the effects of the dark energy sound speed primarily manifest on very large scales where cosmic variance and observational uncertainties remain significant.

Importantly, allowing $c_s^2$ to vary does not noticeably affect the projected constraints on the remaining cosmological parameters. In particular, the uncertainties on the interaction parameter $Q$ and the DE EoS parameters $(w_0, w_a)$ remain essentially unchanged. This indicates that fixing $c_s^2 = 1$ in the baseline analysis does not introduce any significant bias in the forecasted constraints on the iDMDE parameters.

\subsection{Alternative Dark Energy Parametrization: JBP}

The analysis thus far had employed the CPL parametrization of the DE EoS. Although CPL provides a widely used and flexible description of dynamical dark energy, it is useful to verify whether the projected constraints obtained in this work depend sensitively on this specific parametrization. To this end, the analysis was repeated using the JBP parametrization $w(a) = w_0 + w_a a(1-a)$ \citep{Jassal:2005qc}.

Table \ref{tab:JBP_constraints} and Fig. \ref{fig:JBP} show the projected parameter constraints obtained for the interacting JBP parametrization using the combined SKA galaxy clustering and cosmic shear observations in the phantom and non-phantom regimes, respectively. In each case, the SKA forecasts are compared with the baseline constraints derived from current observations (Planck 2018 + DESI DR2 BAO + Pantheon+).

The projected constraints exhibit behaviour similar to that observed for the i-CPL model. In particular, the inclusion of SKA observations leads to a substantial reduction in the allowed parameter space for the interaction parameter $Q$ and the DE EoS parameters. The overall structure of the allowed parameter regions remains comparable between the two parametrizations. These results indicate that the projected improvements from future SKA observations reported in this work are not specific to a single choice of dark energy parametrization. While a more practical treatment of observational systematics and noise may affect the quantitative estimates, the analysis here consistently suggests a significant improvement in precision from those upcoming surveys, compared to current constraints.

\begin{table}
    \resizebox{0.5\textwidth}{!}{\renewcommand{\arraystretch}{1.35} \setlength{\tabcolsep}{20 pt}
    \begin{tabular}{c c c}
        \hline
        \textbf{Parameter} & \textbf{i-JBP Phantom} & \textbf{i-JBP Non-Phantom} \\ 
        \hline
        {\boldmath${100\Omega_b}{h^2}$} & $2.250\pm 0.012            $ & $2.255\pm 0.013            $\\
        {\boldmath${\Omega_c}{h^2}$} & $0.16364\pm 0.00086        $ & $0.14499\pm 0.00080        $\\
        {\boldmath$100{\theta_s}$} & $1.04208\pm 0.00028        $ & $1.04210\pm 0.00029        $\\
        {\boldmath${\ln{\left({10^{10}A_s}\right)}}$} & $3.053\pm 0.013            $ & $3.050\pm 0.013            $\\
        {\boldmath$n_s$          } & $0.9708\pm 0.0021          $ & $0.9733\pm 0.0022          $\\
        {\boldmath${\tau}$       } & $0.0600\pm 0.0068          $ & $0.0590\pm 0.0069          $\\
        {\boldmath$Q$            } & $0.487\pm 0.011            $ & $0.2716\pm 0.0082          $\\
        {\boldmath$w_0$          } & $> -1.02                   $ & $-0.889\pm 0.015           $\\
        {\boldmath$w_a$          } & $-0.771\pm 0.087           $ & $-1.02\pm 0.11             $\\
        \hline
        {\boldmath$H_0$          } & $67.64\pm 0.15             $ & $67.40\pm 0.18             $\\
        {\boldmath$\Omega_{m0}$  } & $0.4083\pm 0.0031          $ & $0.3702\pm 0.0032          $\\
        {\boldmath$\sigma_{8,0}$ } & $0.6323\pm 0.0035          $ & $0.6843\pm 0.0037          $\\
        {\boldmath$S_8$          } & $0.7377\pm 0.0023          $ & $0.7602\pm 0.0022          $\\
        \hline
    \end{tabular}
    }
    \caption{Projected constraints from the combined analysis of SKA1 galaxy clustering and cosmic shear observations, with current CMB+BAO+SNIa data, for the i-JBP phantom and non-phantom models, under the realistic treatment of non-linear scales.}
    \label{tab:JBP_constraints}
\end{table}


\section{Conclusions}\label{sec:conclusion}

Future 21-cm intensity mapping and large-scale galaxy surveys will provide powerful tests of the physics of the dark sector and of other possible extensions to the standard cosmological model. This work investigates the expected sensitivity of upcoming post-reionization observations to interacting dark matter dark energy scenarios with dynamical dark energy. The forecasts here focus on measurements from the SKA-mid and the Euclid mission, using several complementary probes of large-scale structure.

An interacting dark matter dark energy model is considered here, in which dark energy, described by a dynamical equation of state, exchanges energy with cold dark matter through a coupling proportional to the dark energy density. The parameter space therefore includes the six standard cosmological parameters together with the interaction strength $Q$ and the dark energy equation of state parameters $(w_0,w_a)$. Fiducial values were obtained from current constraints derived from Planck 2018 CMB observations combined with DESI DR2 BAO measurements and the Pantheon+ supernovae compilation.

Using mock observations for 21-cm intensity mapping, galaxy clustering and cosmic shear, the constraining power of future SKA-mid surveys was explored, with both conservative and optimistic treatments of theoretical uncertainties at non-linear scales. The results indicate that SKA-mid observations can substantially improve current constraints on interacting dark matter dark energy models. In particular, galaxy clustering and cosmic shear measurements provide the strongest gains in sensitivity, while 21-cm intensity mapping also contributes significant additional information. When the different SKA-mid probes are combined with current data, the projected uncertainties on the interaction parameter $Q$ and the dark energy equation of state parameters $(w_0,w_a)$ are reduced dramatically compared to current limits.

In addition to SKA forecasts, the constraining power of Euclid galaxy clustering and cosmic shear measurements was also assessed. Euclid is expected to offer significant improvements relative to current CMB+BAO+SNIa data as an important complementary probe of the dark sector. However, SKA generally yields stronger projected constraints, primarily because it can probe the large-scale matter distribution across a wider redshift range through both radio galaxy surveys and 21-cm intensity mapping.

The robustness of the forecasts were examined under several modelling assumptions. Repeating the analysis in the non-phantom regime yielded results broadly similar to those obtained for the phantom case. Allowing the dark energy sound speed to vary does not significantly affect the constraints on the other cosmological parameters, and the data remain largely insensitive to this quantity. Likewise, adopting an alternative parametrization of the dark energy equation of state, such as the JBP form, leads to results consistent with those obtained using the CPL model. This indicates that the forecasted improvements from future surveys are not strongly dependent on the specific parametrization adopted for dynamical dark energy.

Overall, the results highlight the considerable potential of upcoming large-scale structure and 21-cm surveys to test and constrain possible interactions within the dark sector and to probe dynamical dark energy models. The combination of SKA and Euclid observations will provide highly complementary measurements of the large-scale matter distribution across cosmic time, offering a promising opportunity to explore possible non-gravitational interactions between dark matter and dark energy.


\section*{Acknowledgements}

We thank Debarun Paul for insightful discussions. We also thank the anonymous reviewer for their valuable suggestions, which helped improve the manuscript. RS and AD thank the Indian Statistical Institute (ISI) Kolkata for financial support through Senior Research Fellowship. PM acknowledges the Anusandhan National Research Foundation (ANRF), Government of India, for financial support under the National Post-Doctoral Fellowship (N-PDF File no. PDF/2023/001986). SP thanks ANRF,  Government of India, for partial support through Project No. CRG/2023/003984. We acknowledge the computational facilities provided by the SyMeC high performance computing (HPC) cluster of ISI Kolkata and the Pegasus HPC facility at the Inter-University Centre for Astronomy and Astrophysics (IUCAA), Pune, India.


\section*{Data Availability}

The datasets used in this work are all publicly available. The modified codes used for this study may be made available upon reasonable request.


\appendix

\section{Constraints on interacting CPL and interacting JBP from current datasets}\label{appendix}

The mean and 1$\sigma$ constraints obtained for i-CPL and i-JBP using current datasets are presented in Table \ref{tab:current} (reproduced from \citet{Shah:2025ayl}).

\begin{table}
    \resizebox{0.5\textwidth}{!}{\renewcommand{\arraystretch}{1.5} \setlength{\tabcolsep}{3 pt}
    \begin{tabular}{c c c c c}
        \hline
        \textbf{Parameters} & \textbf{i-CPL Phantom} & \textbf{i-CPL Non-Phantom} & \textbf{i-JBP Phantom} & \textbf{i-JBP Non-Phantom} \\ 
        \hline
        {\boldmath${100\Omega_b}{h^2}$} & $2.247\pm 0.013 $ & $2.254\pm 0.013 $ & $2.252\pm 0.013 $ & $2.248\pm 0.013 $\\
        {\boldmath${\Omega_c}{h^2}$} & $0.1556^{+0.0086}_{-0.0071}$ & $0.048^{+0.025}_{-0.032} $ & $0.151^{+0.013}_{-0.010} $ & $0.108^{+0.042}_{-0.034} $\\
        {\boldmath$100{\theta_s}$} & $1.04204\pm 0.00028 $ & $1.04210\pm 0.00027 $ & $1.04209\pm 0.00026 $ & $1.04206\pm 0.00028 $\\
        {\boldmath${\ln{\left({10^{10}A_s}\right)}}$} & $3.047\pm 0.015 $ & $3.051\pm 0.016 $ & $3.052\pm 0.016 $ & $3.048\pm 0.015 $\\
        {\boldmath$n_s$ } & $0.9685\pm 0.0036 $ & $0.9710\pm 0.0034 $ & $0.9701\pm 0.0037 $ & $0.9690\pm 0.0037 $\\
        {\boldmath${\tau}$ } & $0.0565\pm 0.0072 $ & $0.0593\pm 0.0078 $ & $0.0599\pm 0.0083 $ & $0.0575\pm 0.0074 $\\
        {\boldmath$Q$ } & $0.387^{+0.099}_{-0.084} $ & $-0.52^{+0.16}_{-0.21} $ & $0.34^{+0.14}_{-0.11} $ & $-0.07\pm 0.30 $\\
        {\boldmath$w_0$ } & $> -1.04 $ & $-0.761^{+0.090}_{-0.073} $ & $> -1.06 $ & $-0.783^{+0.097}_{-0.12} $\\
        {\boldmath$w_a$ } & $-0.33\pm 0.14 $ & $-0.188^{+0.088}_{-0.10} $ & $-0.35\pm 0.30 $ & $-1.13\pm 0.43 $\\
        \hline
        {\boldmath$H_0$ } & $67.94\pm 0.53 $ & $67.67\pm 0.61 $ & $68.07\pm 0.54 $ & $67.62\pm 0.58 $\\
        {\boldmath$\Omega_{m0}$ } & $0.387^{+0.021}_{-0.019} $ & $0.155^{+0.055}_{-0.069} $ & $0.375^{+0.029}_{-0.023} $ & $0.286^{+0.093}_{-0.075} $\\
        {\boldmath$\sigma_{8,0}$ } & $0.661^{+0.025}_{-0.030} $ & $1.73^{+0.59}_{-0.75} $ & $0.674^{+0.035}_{-0.047} $ & $0.97^{+0.24}_{-0.47} $\\
        {\boldmath$S_8$ } & $0.751^{+0.014}_{-0.015} $ & $1.15^{+0.20}_{-0.23} $ & $0.752^{+0.018}_{-0.022} $ & $0.878^{+0.098}_{-0.17} $\\
        \hline
    \end{tabular}
    }
    \caption{The mean and 1$\sigma$ constraints obtained for i-CPL and i-JBP using combined Planck 2018 + DESI DR2 BAO + Pantheon+ datasets.}
    \label{tab:current}
\end{table}


\bibliographystyle{mnras}
\bibliography{mnras} 

\bsp
\label{lastpage}
\end{document}